\documentclass[apj]{emulateapj}

\usepackage{epsf}
\usepackage{psfig}
\usepackage{graphicx}
\usepackage{natbib}
\usepackage{enumerate}
\usepackage{float}
\usepackage{subfloat}

\begin{document}
\title{Disk-Loss and Disk-Renewal Phases in Classical Be Stars. II. Contrasting with Stable and Variable Disks}
\author{Zachary H. Draper\altaffilmark{1,2}, John P. Wisniewski\altaffilmark{3}, Karen S. Bjorkman\altaffilmark{4}, Marilyn R. Meade\altaffilmark{5}, Xavier Haubois\altaffilmark{6,7}, Bruno C. Mota\altaffilmark{6}, Alex C. Carciofi\altaffilmark{6},  Jon E. Bjorkman\altaffilmark{4}}

\altaffiltext{1}{Department of Physics and Astronomy, University of Victoria, 3800 Finnerty Rd, Victoria, BC V8P 5C2 Canada}
\altaffiltext{2}{Herzberg Institute of Astrophysics, National Research Council of Canada, Victoria, BC V9E 2E7 Canada}
\altaffiltext{3}{HL Dodge Department of Physics \& Astronomy, University of Oklahoma, 440 W Brooks St, Norman, OK 73019 USA, wisniewski@ou.edu}  
\altaffiltext{4}{Ritter Observatory, Department of Physics \& Astronomy, Mail Stop 113, University of Toledo, Toledo, OH 43606 USA, karen.bjorkman@utoledo.edu, jon@physics.utoledo.edu}
\altaffiltext{5}{Space Astronomy Lab, University of Wisconsin-Madison, 1150 University Avenue, Madison, WI 53706 USA, meade@astro.wisc.edu}
\altaffiltext{6}{Instituto de Astronomia, Geof\'{i}sica e Ci\^{e}ncias Atmosf\'{e}ricas, Universit\'{a}ria de S\~{a}o Paulo, Rua do Mat\~{a}o 1226, Cidade Universit\'{a}ria, 05508-900 S\~{a}o Paulo, SP Brazil, xhaubois@astro.iag.usp.br, carciofi@usp.br}
\altaffiltext{7}{Sydney Institute for Astronomy, School of Physics, University of Sydney, NSW 2006, Australia}

\begin{abstract} 
Recent observational and theoretical studies of classical Be stars have established the utility of polarization color diagrams (PCD) in helping to 
constrain the time-dependent mass decretion rates of these systems.  We expand on our pilot observational study of this phenomenon, and report 
the detailed analysis of a long-term (1989-2004) spectropolarimetric survey of 9 additional classical Be stars, including systems exhibiting evidence of partial disk-loss/disk-growth episodes as well as systems exhibiting long-term stable disks.   After carefully characterizing and removing the interstellar 
polarization along the line of sight to each of these targets, we analyze their intrinsic polarization behavior.  We find that many steady-state Be disks pause
at the top of the PCD, as predicted by theory.  We also observe sharp declines in the Balmer jump polarization for later spectral type, near edge-on steady-state disks, again as recently predicted by theory, likely caused when the base density of the disk is very high, and the outer region of the edge-on disk starts to self absorb a significant number of Balmer jump photons.  The intrinsic $V$-band polarization and polarization position angle of $\gamma$ Cas exhibits variations that seem to phase with the orbital period of a known one-armed density structure in this disk, similar to the theoretical predictions of Halonen \& Jones.  We also observe stochastic jumps in the intrinsic polarization across the Balmer jump of several known Be+sdO systems, and 
speculate that the thermal inflation of part of the outer region of these disks could be responsible for producing this observational phenomenon.  Finally, we estimate the base densities of this sample of stars to be between $\approx 8\times 10^{-11}$ to $\approx 4 \times 10^{-12}\,\rm g cm^{-3}$ during quasi steady state periods given there maximum observed polarization.

\end{abstract}


\section{Introduction} \label{intro}

Classical Be stars are a subset of B-type main sequence stars which are characterized by their rapid rotational velocities ranging from 60\% to 100\% of their critical rate \citep{riv13}.  They have a geometrically flattened decretion disk that is fed from material from the stellar photosphere 
 as diagnosed from studies of their optical/IR emission lines, polarization, and interferometric signatures (see e.g. \citealt{por03,stee11}).  A large volume of observations suggest the kinematic properties of these gas disks is best represented by near Keplerian rotation \citep{hum00, me07a, pot10, whe12, kra12}.  For the most up to date review of Classical Be stars see \cite{riv13}.

As summarized in \citet{car11}, the viscous decretion disk model developed by \citet{lee91} can explain many of the observational signatures of Be disks, although other models have been explored to explain the structure of these disks \citep{bjo93,cas02,bro08}.  One key unanswered question in the study 
of Be disks is what mechanism(s) are responsible for injecting material into these disks.  Non-radial pulsations have been suggested to be one contributing factor to supplying material to some of these disks \citep{cra09,riv98,nei02}, while periastron passage of binary companions may contribute 
in other systems such as $\delta$ Scorpii \citep{mir01,mir03}.  This scenario for $\delta$ Scorpii is now questionable given the disk's growth prior to the periastron passage of 2011 \citep{miro13}.  Nonetheless, binarity may play a role in the phenomenon and disk variability.  For example, the source of material and angular momentum in non-classical Be stars can be the result of a red giant phase binary transferring material to create a Be+sdO system \cite{gies98}. Characterizing the evolution of Be stars' mass-loss rates
is another promising approach to constrain the disk-feeding mechanism.  On short time-scales, \cite{car07} noted polarimetric variability in 
Archernar likely arising from injections of discrete blobs of mass into the inner disk, that subsequently circularize into rings.  Studying longer duration 
disk-loss and disk-regeneration events \citep{und82,doa83,cla03,vin06,hau12}, including the time-scales \citep{wis10} and statistical frequency \citep{mcs08,mcs09} of these episodes, is another way to diagnose the mechanism feeding Be disks.  
 
Polarimetry has been used to study the Be phenomenon for both individual Be stars \citep{qui97,woo97,cla98} and larger statistical surveys \citep{coy69,mcl78,poe79,wi07b}.  It is widely believed that Thompson scattering (free-electron scattering) is the source of polarization in Be stars \citep{wo96a,wo96b,hal13}.  Pre- or post-scattering absorption of photons within the disk can imprint wavelength dependent signature on top of the wavelength independent Thompson scattering \citep{woo95}.  Because the polarization across the Balmer jump traces material in the innermost regions of disks ($\sim$6 R$_{\star}$; \citealt{car11}, \citealt{ha13a}) and the $V$-band polarization is a tracer of the total scattering mass of the disk, studying the time evolution of  
the wavelength-dependence of polarization in Be stars can be used to constrain the time dependence of the mass decretion rate and the $\alpha$ parameter in these systems.  Specifically, the slope, shape, and temporal evolution of polarization color diagrams (PCD) observed in Be systems \citep{dra11} have been theoretically reproduced, for the first time, by time-dependent radiative transfer models in which the mass decretion rate, $\alpha$ parameter, and inclination angle are the primary variables \citep{hau13}.  In addition to density changes caused by changes in the mass decretion rate, it has been suggested that PCD diagram loops can also be produced by one-armed density perturbations \citep{ha13b}.

In this paper, we implement the PCD diagram diagnostic developed in \citet{dra11} on a broader sample of Be stars, including systems showing evidence of experiencing disk-loss events and those whose disks appear roughly stable over time.  We describe the data sample in Section 2, and discuss the techniques we used to remove the interstellar polarization component from each dataset in Section 3.  We discuss the temporal evolution of each Be disk in PCD diagram parameter space in Section 4, and also detail evidence of variability in the disk position angle in select systems.   Finally, we summary the major results of this manuscript in Section 5.

\section{Observations and Data Reduction} \label{data}

The spectropolarimetric data analyzed in this study were obtained by the University of
Wisconsin's (UW) HPOL spectropolarimeter, mounted on the 0.9m 
Pine Bluff Observatory (PBO) telescope.  
Data obtained before 1995 were recorded using a dual Reticon array detector spanning the wavelength
range of 3200-7600\AA\, with a spectral resolution of 25\AA\    
 \citep{wol96}.  Beginning in 1995, HPOL's detector was upgraded to a 400 x 1200 pixel CCD camera that
 provided coverage from 3200-6020\AA\ at a resolution of 10\AA\ and 5980-10,500\AA\ 
 at a resolution of 7\AA\ \citep{nor96}.
Further details about HPOL can be found in \citet{noo90}, \citet{wol96}, and \citet{har00}.  

Data obtained by HPOL were reduced and calibrated using REDUCE, a spectropolarimetric software package developed by the University of Wisconsin-Madison (see \citealt{wol96}).  
Routine monitoring of unpolarized standard stars at PBO has enabled the instrumental polarization to be carefully calibrated.
The residual instrumental systematic errors depend mildly on the date of the observations, but range from 0.027-0.095\% in the U-band, 0.005-0.020\% in the $V$-band, and 0.007-0.022\% in the I-band.
At the current time, HPOL data from 1989 to 2000 are available on the STScI MAST archive \footnote{See archive website for more up to date details: http://archive.stsci.edu/hpol/}.

HPOL spectroscopic data were not flux calibrated to an absolute level because of nonphotometric skies typically present during the observations.
To mitigate any relative flux offsets between red and blue grating data obtained on the same night, we applied a constant multiplicative factor to the 
grating with the lower flux in an observation, and then merged the red and blue grating bandpasses. 

When available, HPOL observations were supplemented with observations from the Wisconsin Ultraviolet Photo-Polarimeter Experiment (WUPPE) from both the Astro-1 and Astro-2 missions flown on the Space Shuttles Columbia and Endeavor, respectively.
WUPPE is a 0.5 meter telescope with a spectropolarimeter which simultaneously obtained spectra and polarization from 1500 to 3200\AA\ with a resolution of about 16\AA\ .
For a more detailed explanation of the instrument see \cite{nord94}, and \cite{clay97} for a detailed description of the pre-flight and in-flight calibrations.

Additional spectroscopic observations of 66 Oph obtained with the fiber-fed echelle spectrograph mounted on the Ritter Observatory 1m telescope were also analyzed.  These observations were obtained with the default spectrograph setup, yielding R $\sim$26,000.  These data were reduced using standard IRAF techniques.  A summary of the data used for each star in this study is in Table 1. 

\section{Data Analysis}

\subsection{Interstellar Polarization} \label{isp}

Before any robust interpretations of polarization from our classical Be stars can be made, the interstellar polarization (hereafter, ISP) must be removed.  
The ISP is due to a dichroic scattering in the interstellar medium (hereafter, ISM) imprinting itself on the intrinsic polarization of the science target.  
There are three common techniques which are used to characterize the ISP: field stars, emission lines, and utilizing the wavelength dependence of the observed polarization. \citep{wi07a}.
In this paper, we take advantage of the field star and wavelength dependence to diagnose the ISP for our target stars and try to present an improved and streamlined process for determining ISP for spectropolarimetry from long surveys, like HPOL.

\subsubsection{48 Lib} \label{48libpol}

Using our IDL routines, the calibrated HPOL data was processed in a manner similar to the ISP removal process in \cite{qui97} and \cite{wis10} to constitute a standardized ISP determination process.
We measure the raw Johnson $V$-band polarization measurements from the spectropolarimetry using equations \ref{eq:filter} and \ref{eq:filtererr}, 

\begin{equation} 
\label{eq:filter}
\%P = \int_{\lambda_{1}}^{\lambda_{2}} \frac{P(\lambda)*F(\lambda)*w(\lambda)}{F(\lambda)*w(\lambda)}
\end{equation}

\begin{equation} 
\label{eq:filtererr}
\%P_{err} = \frac{1}{\sqrt{n}} \int_{\lambda_{1}}^{\lambda_{2}} d\lambda \frac{\rm{Err}(\lambda)*F(\lambda)*w(\lambda)}{F(\lambda)*w(\lambda)} 
\end{equation}

\noindent
where P($\lambda$) is a stokes parameter of the spectropolarimetry ($Q$, $U$, or $P$), F($\lambda$) is the relative flux, and w($\lambda$) is the filter function desired to weight the data.
The error measurement is divided by a Poisson statistic, where $n$ is the total number of data points within the wavelength range.  This is consistent with the \textit{pfil} command in the REDUCE software code developed for HPOL (\citealt{wo96a}).  These raw $V$-band polarization data are compiled in Table \ref{ip_data_raw} for every target in our sample.

When they are not spatially resolved, Be systems exhibit linear polarization along a single, preferred position angle, defined by the orientation of the 
disk's major axis on the sky.  We fit a linear trend through the data in the $QU$ plane to determine this mean disk polarization axis 
(Fig \ref{fig:isp_all}, left column).
Then we rotated the entire data set by the mean intrinsic disk angle, thereby putting them into a rotated ($Q$' $U$') space.  The ISP perpendicular component ($P_{\perp}$) to the intrinsic polarization, i.e. along the $U$' axis, in this regime is independent of the intrinsic disk polarization.
The $Q$' axis is comprised of the parallel component of ISP ($P_{\parallel}$) plus the intrinsic disk polarization.

The $U$' axis for all observations were then combined with a error weighted mean (Fig \ref{fig:isp_all}, middle column). 
To extend the wavelength coverage of these data, which aids our effort to robustly model the wavelength dependent ISP signature described below, 
we supplemented our target stars with archival WUPPE data where possible (Table \ref{star_sum}).  These archival UV data were rotated by the same intrinsic 
disk angle determined for each star and merged with the optical $U$' spectropolarimetry.  We then fit these $U$' data with a modified version of the empirical
Serkowski law \citep{ser75,wil82}, 

\begin{equation}
P(\lambda) = P_{\perp} \exp \left [-K \ln ^{2} (\lambda _{\rm max}/\lambda) \right ]
\end{equation}

\begin{equation}
K = (1.68*10^{-4})*(\lambda_{\rm max})-0.002
\end{equation}

\noindent
to determine the perpendicular component of ISP.  $\lambda_{\rm max}$ can be determined by the wavelength of the Serkowski curves' inflection point at the peak amplitude of ISP. This characteristic Serkowski profile stems from dichroic absorption of the ISM where certain wavelengths are preferentially scattered from dust grains aligned with the galactic magnetic field.  

Following the methodology used in \citet{qui97}, we compute the ISP$_{\parallel}$ component by assuming that the PA of the total ISP is sufficiently represented by nearby field stars to each of our targets, and then using simple geometry to extract the magnitude of the ISP$_{\parallel}$ component.  
For each of our science targets, we generated a field star PA estimate ($\rm PA_{fs}$) by selecting stars from the \citet{hei00} catalog in the vicinity of each science target (3 degree maximum separation and within 150 pc of the target) that exhibit no conspicuous intrinsic polarization (Table \ref{t2}).  We separated the $V$-band polarization for each field star into $Q$ and $U$ components, computed an error weighted mean, and finally used this error weighted mean to compute the final field star PA for each science target (Table \ref{t2}).  The error in this PA was computed using the standard deviation of $QU$:

\begin{equation}
\rm PA = 0.5 * \tan^{-1} \left( \frac {\textit{U}} {\textit{Q}} \right)
\end{equation}

\begin{equation}
\label{paerr}
\rm PA_{err} = \frac {\sigma \textit{U} * \textit{Q} + \sigma \textit{Q} * \textit{U}} {\textit{U}^{2}+\textit{Q}^{2}}
\end{equation}

\noindent
Given the amplitude of ISP$_{\perp}$ and the field star PA, we compute the amplitude of ISP$_{\parallel}$ using simple geometry:

\begin{equation}
P_{\parallel} = \frac {P_{\perp}} {\tan (2*\rm{PA_{fs}})}
\end{equation}

\noindent
Finally, the total ISP is computed using the amplitude of ISP$_{\perp}$ and ISP$_{\parallel}$: 
\begin{equation}
P_{\rm ISP} = \sqrt{P_{\perp}^2 + P_{\parallel}^2}
\end{equation}

\noindent
With the $P_{\rm ISP}$, $\lambda_{\rm max}$ and $\rm{PA_{fs}}$ the ISP characterization is complete.  As a check of the accuracy of our ISP removal, we plot the resultant intrinsic polarization on a Stokes QU diagram and confirm that the data pass through the origin (Fig. 1A, B, and C, right column) with 48 Lib shown in Fig. 1A. The final ISP values, along with the parallel and perpendicular components, are tabulated in Table \ref{isp} for each science star.  For 48 Lib, our final ISP parameters of $P_{\rm max}$ = 0.86\% and $\theta$ = 93$^{\circ}$ are consistent with the ISP PA derived from previous field star (82 $\pm$ 22) and H$\alpha$ line depolarization (85 $\pm$ 3) studies \citep{poe79} and amplitude of ISP derived from these studies (0.64\% - 0.73\%; \citealt{poe79}).  Removing this ISP component yields the time-dependent intrinsic polarization behavior within our data (Fig \ref{fig:ip_all}), which is compiled in Table \ref{ip_data} for all targets in our sample.

\subsubsection{$\phi$ Per} \label{phiperpol}

When the intrinsic polarization angle was not well determined, the ISP removal process described in Section \ref{48libpol} failed, as diagnosed by the combined $U$' data set exhibiting a Balmer jump instead of a characteristic Serkowski curve.  For $\phi$ Per this breakdown occured because the source exhibited minimal variability in its intrinsic polarization levels, such that the raw data were not well fit by a linear trend on the $QU$ diagram (see Fig \ref{fig:isp_all_a}).  To mitigate this, instead of using the $V$-band polarization to determine the intrinsic PA on the $QU$ diagram we used the raw polarization data across the Balmer jump 3200-4000 $\rm{\AA}$.  The differing opacity across the Balmer jump imprints a change in amplitude of linear polarization across the jump.  Plotting the polarization across the Balmer jump in $QU$ space helps to differentiate the intrinsic polarization of the disk, thereby allowing us to fit these data with a linear trend to derive the intrinsic disk position angle.  Once the intrinsic disk PA was determined, we adopted the ISP PA of \citet{qui97} and determined the ISP parameters using the rest of the procedures described in Section 3.1.1.  Our total ISP polarization, $P_{\rm max}$ = 0.76\%, was similar to that found by \citet{qui97}, 0.82\%.  The resultant time-dependent intrinsic polarization of $\phi$ Per is shown in 
Fig \ref{fig:ip_all_a} and Table \ref{ip_data}.

\subsubsection{$\gamma$ Cas} \label{gammacaspol}

Like $\phi$ Per, the long-term stability of $\gamma$ Cas's disk necessitated use of its polarization across the Balmer jump to characterize the intrinsic disk position angle (Fig \ref{fig:isp_all_a}).   We adopted the field star PA used by \citet{mcl78} and \citet{qui97} (see Table \ref{isp}).  The final ISP parameters we derived, $P_{\rm max}$ = 0.31\% and $\theta$ = 95$^{\circ}$ (Table \ref{isp}), were consistent with those derived by \citet{qui97}, $P_{\rm max}$ = 0.26\% and $\theta$ = 95$^{\circ}$.  The time-dependent intrinsic polarization of $\gamma$ Cas is shown in Fig \ref{fig:ip_all_a} and Table \ref{ip_data}.

\subsubsection{66 Oph} \label{66ophpol}

We adopted the same ISP method used for 48 Lib on 66 Oph.  After defining the major axis of the disk by fitting a linear regression to the $V$-band polarization data plotted in a QU diagram (Fig \ref{fig:isp_all_b}), we determined ISP$_{\perp}$ by fitting a Serkowski law to the U' data and ISP$_{\parallel}$ from the PA of nearby field stars (Table \ref{t2}).  The final ISP parameters we derived, P$_{max}$ = 0.45\% and $\theta$ = 81$^{\circ}$ (Table \ref{isp}), were consistent with those derived by \citet{poe79} from field star and H$\alpha$ line depolarization, $P_{\rm max}$ = 0.51-0.52\% and $\theta$ = 82$\pm$3 - 86$\pm$19.  The time-dependent intrinsic polarization of 66 Oph is shown in Fig \ref{fig:ip_all_b} and Table \ref{ip_data}.

\subsubsection{$\omega$ Ori} \label{omegaoripol}

$\omega$ Ori exhibited significant $V$-band polarization variability, which enabled us to determine the PA of the disk's major axis by fitting a linear regression to these data in the Stokes QU diagram (Fig \ref{fig:isp_all_b}).  While we were able to determine ISP$_{\perp}$ by fitting a Serkowski law to the U' data (middle panel, Fig \ref{fig:isp_all_b}; Table \ref{isp}), the field star PA estimates for this source varied from 20-165 (Table \ref{t2}), with an average value of 36$\pm$9.7$^{\circ}$.  Previous estimates of the ISP towards $\omega$ Ori indicated that the magnitude of ISP was low ($<$0.05\%, \citealt{poe79} to $\sim$0.1\%, \citealt{mcl78}), and the orientation of this ISP ranged from 151$^{\circ}$ \citep{poe79} to being undetermined \citep{mcl78}.  Our long temporal baseline of spectropolarimetric data is particularlly helpful in constraining the ISP PA as Figs. \ref{fig:isp_all_b} and \ref{fig:ip_all_b} indicate that $\omega$ Ori experienced disk depletion events.  Because the variation in the raw $V$-band polarization (Fig \ref{fig:ip_all}) passes near the origin in QU space, this excludes our average field star PA (36$^{\circ}$) from being a viable ISP PA.  Furthermore, an adopted ISP PA of 36$^{\circ}$ would cause the $V$-band polarization and H$\alpha$ to have an anti-correlated trend with time.  Rather the total ISP PA must lay in the third quadrant of the QU diagram (90-135$^{\circ}$).  We estimate this PA to be $\sim$110$^{\circ}$, under the assumption that our lowest polarization data captures the system in a near disk-less state, akin to the scenario considered by \citet{wis10}.  Removing our estimate of the total ISP, $P_{\rm max}$ = 0.23\% and $\theta$ = 110$^{\circ}$, from these data yields the time-dependent intrinsic polarization shown in 
Fig. \ref{fig:ip_all_b} and Table \ref{ip_data}.  

\subsubsection{$\psi$ Per} \label{psiperpol}

$\psi$ Per exhibited significant variability in its total $V$-band polarization across the time sampling of our data (Fig \ref{fig:isp_all_b}), enabling us to determine the PA of the disk's major axis.  The perpendicular component of ISP was weak and poorly constrained by a Serkowski law (middle panel, Fig \ref{fig:isp_all_b}).   We adopted a field star PA of 112$^{\circ}$, yielding final ISP parameters of $P_{\rm max}$ = 0.09\% and $\theta$ = 112$^{\circ}$.  These parameters are generally consistent with those determine by \citet{poe79} ($P_{\rm max}$ = 0.31$\pm$0.11\% - 0.36$\pm$0.10\%; $\theta$ = 128$\pm$13 - 129$\pm$10) and \citet{mcl78} ($P_{\rm max}$ = 0.4\% and $\theta$ = 135$^{\circ}$).  However, they are finding a stronger parallel ISP component. Removing this ISP yields the time-dependent intrinsic polarization shown in Fig. \ref{fig:ip_all_b} and Table \ref{ip_data}.

\subsubsection{28 Cyg} \label{28cygpol}

We determined the PA of the disk major axis of 28 Cyg by fitting the variation in its $V$-band polarization in the QU frame with a linear regression 
(Fig. \ref{fig:isp_all_c}).  Previous studies have suggested a weak ISP along the line of sight to 28 Cyg ($<$0.20\%, \citealt{poe79}; 0.1\%, \citealt{mcl78}), which agrees with our new analysis.  We found 28 Cyg had a very weak ISP$_{\perp}$ component (Fig \ref{fig:isp_all}; Table \ref{isp}).  Based on a field star PA of 79$^{\circ}$ (Table \ref{t2}), we determined final ISP parameters of $P_{\rm max}$ = 0.25\% and $\theta$ = 79$^{\circ}$.
Removing this small ISP component yielded the time-dependent intrinsic polarization shown in Fig. \ref{fig:ip_all_c} and Table \ref{ip_data}.

\subsubsection{FY CMa} \label{fycmapol}

Both the raw $V$-band polarization (Fig \ref{fig:isp_all_c}) and polarization across the Balmer jump were well fit with a linear regression passing near the origin.  This unique situation presents substantial challenges for determining the ISP, as it either implies that the ISP component is zero or it implies that the ISP component is aligned with the intrinsic disk axis.  In the latter scenario, one would only be able to robustly determine the ISP during epochs in which FY CMa experienced a complete loss of its disk (see e.g. \citealt{wis10}).  For the purposes of this paper we have assumed the ISP component is small ($P_{\rm max}$ = 0.04\% and $\theta$ = 136$^{\circ}$).  The nominal field star PA, 136$^{\circ}$ (Table \ref{t2}), is $\sim$20$^{\circ}$ from the PA of the disk major axis, which supports our assumption that the ISP towards FY CMa is not parallel with the intrinsic PA.  The time-dependent 
intrinsic polarization for FY CMa is shown in Fig. \ref{fig:ip_all_c} and Table \ref{ip_data}.

\subsubsection{59 Cyg} \label{59cygpol}

Like $\phi$ Per and $\gamma$ Cas, the long-term stability of 59 Cyg's $V$-band polarization (see Fig \ref{fig:isp_all_c}) 
forced us to measure the change in polarization across the Balmer jump to characterize the intrinsic disk position angle.  We were able to 
determine ISP$_{\perp}$ by fitting a Serkowski law to the UV and optical U' data (Fig. \ref{fig:isp_all_c}); however, the total ISP PA suggested by field 
stars surrounding 59 Cyg (29 $\pm$35$^{\circ}$; Table \ref{t2}) is unlikely to be correct as this lays parallel to the intrinsic disk PA.  This could only be plausible if the ISP$_{\perp}$ was instead zero.  We therefore were 
unable to determine a robust total ISP for 59 Cyg.  As will be discussed in Section \ref{sdO}, we did explore the behavior of 59 Cyg's mostly intrinsic polarization across the Balmer jump (Fig. \ref{fig:ip_all_c}) assuming a total ISP around 0$^{\circ}$.  Specifically, we explored ISP parameter space about this adopted value and confirmed that, even though there is significant uncertainty in the correct total ISP PA, this did not specifically affect the observed 
Balmer jump polarization discussed in that section.

\subsubsection{60 Cyg and $\pi$ Aqr} \label{60cygpol}

\citet{wis10} already used the total disk loss phases of 60 Cyg and $\pi$ Aqr to establish well determined ISP parameters for these systems.  
We applied the IDL code and basic techniques used throughout this paper on these data (see Figs. \ref{fig:isp_all} and \ref{fig:ip_all}), 
and adopted the same total ISP PA as used in \citet{wis10}.  This enabled us to 
test the effects of using different binning during the fitting process as compared to the earlier work (e.g. 60 Cyg) as well as the effects of including 
UV polarimetry and different binning (e.g. $\pi$ Aqr).  As seen in Table \ref{isp}, the ISP parameters we derive are fully consistent with those reported 
in \citet{wis10}.  

\subsection{H$\alpha$ Equivalent Widths} \label{isp}

To help assess the temporal behavior of the disk systems in our sample, we computed H$\alpha$ EWs from our spectropolarimetric data in the last column of Table \ref{ip_data}.  We processed HPOL flux data using IDL code that computed EWs using the trapezoid rule:

\begin{equation}
EW = \displaystyle\sum_{n=\lambda_{1}}^{\lambda_{2}} \left( 1- \frac{0.5*(F(n)_{line}+F(n+1)_{line})}{F_{cont}} \right) \Delta\lambda
\end{equation}

where $F_{cont}$ is the continuum flux computed from 6366-6456 \AA\ and 6666-6756 \AA. The H$\alpha$ line flux 
was computed from 6516-6606\AA\ ($\lambda_{2}$ and $\lambda_{1}$).\\

Minor variations in H$\alpha$ EWs can be seen in HPOL data that span the change from a Reticon to a CCD detector in 1995.  These variations are not considered real, but rather are likely due to the slight change in spectral resolution achieved between the two different instrument configurations (see Section \ref{data}).  A small number of archival spectroscopic observations of 66 Oph (Table \ref{66Oph_rit_data}), obtained at Ritter Observatory, were also analyzed in this paper to supplement the HPOL data.  As noted previously by \citet{wis10}, there is also a flux offset between H$\alpha$ EWs computed from HPOL and Ritter data, likely owing to the different resolutions of these data.  Since we are 
merely interested in using these data to track the long-term evolution of 66 Oph, this offset does not affect our analysis or interpretations.  \\

\section{Discussion} \label{discussion}

\subsection{Additional Disk-Growth and Loss Phases}

In paper I of this series, we analyzed well sampled, long-term spectropolarimetric data for two classical Be stars 
that clearly underwent a complete disk-loss phase.  In this paper, we explore the behavior of a larger sample of classical Be stars, including systems that exhibit disks that are stable over time periods of more than a decade, as well as systems that exhibit at least partial disk-loss/disk-renewal phases.  While detailed modeling of each system that exhibits a 
non-stable disk is clearly necessary and warranted, albeit outside the scope of this paper, we discuss some of the basic 
observational properties of these systems below. 

\subsubsection{$\psi$ Per}

The intrinsic polarization and H$\alpha$ EW of $\psi$ Per both exhibit clear evidence of significant variability over 
the $\sim$13 year time coverage of our data (Fig. \ref{fig:ip_all_b}).  Throughout the early 1990s, the steady increase in the system's intrinsic polarization, lasting for $\sim$1400 days, accompanied by a small strengthening in its H$\alpha$ EW is indicative of a growth in density and size of the disk.  Although the data sampling is sparse, there is evidence that the disk generally stabilized in strength between 1995-2001.  Figure \ref{fig:ip_all_b} also exhibits a dramatic, monotonic drop in intrinsic polarization over a period of $\sim$190 days starting in August 2002, along with a more gradual, time-delayed, significant decrease in H$\alpha$ EW.  These trends are consistent with a major inside-out clearing of significant mass from the disk \citep{wis10}.  Since a small level of both intrinsic polarization and H$\alpha$ emission remained after the event, we characterize this as a incomplete disk-loss event.  

Characterizing the time-scale of disk-loss events is important in order to constrain the viscosity parameter, $\alpha$, assuming the disk dissipates on a viscous time-scale.  Equation (19) of \citet{bjo05} describes such an assumption and can be rewritten to give $\alpha$ as function of the diffusion timescale, $t_{\rm diff}$. 

\begin{equation}
\alpha = (0.2 \rm yr / t_{\rm diff}) * (r/R_{\star})^{0.5}
\end{equation}

Assuming the bulk of the scattering events producing the observed polarization occurs at a radial distance of $\sim$5 
R$_{\star}$ and the (partial) disk-loss time-scale of $\sim$ 190 days yields $\alpha$ $\sim$0.86 for $\psi$ Per.  This value is intermediate compared to estimates for 60 Cyg ($\sim$0.1; \citealt{wis10}) and robustly measured for 28 CMa (1 $\pm$ 0.2; \citealt{car12}).  Because the $\psi$ Per disk did not completely clear out, our quoted disk-loss time-scale is a lower limit and correspondingly the quoted $\alpha$ parameter is an upper limit.  Clearly, characterizing the viscosity parameter for a larger number of systems is required to assess whether there is a preferred parameter for most Be disks.

\subsubsection{$\omega$ Ori}

The overall steady decline in H$\alpha$ EW strength throughout most of the 13 years of coverage in our dataset (Fig. \ref{fig:ip_all_b}) is indicative of a gradual, albeit incomplete, loss of the system's disk.  The $\sim$1500 day duration decline in intrinsic $V$-band polarization starting at the beginning of our dataset is consistent with a gradual loss of the system's disk.  Mirroring the scenario observed in $\pi$ Aqr reported by \citet{wis10}, the subsequent $\sim$1100 day long increase in intrinsic polarization is likely responsible for the temporary halt in the decline of H$\alpha$ EW strength between 1995-1998 and indicative of temporary replenishment of material to the inner regions of the disk.  The cessation 
of this mass injection to the inner disk is marked by a fast decrease in the intrinsic polarization level, as well as a resumption of the gradual decline in H$\alpha$ EW.  Near the end of our dataset, the sharp rise in both intrinsic polarization and H$\alpha$ EW indicate significant mass injection into the disk.  Overall, these disk-loss and growth epochs occurred in the span of 2-4 years.

\subsubsection{66 Oph} The H$\alpha$ EW of 66 Oph also exhibited a steady decline throughout the entire 13 year baseline of 
our dataset (Fig. \ref{fig:ip_all_b}).  Supplementary H$\alpha$ EWs from spectra obtained at Ritter Observatory improve the time sampling during the end of our spectropolarimetric data, and confirm that the emission strength decline continues during these epochs.  The Ritter EWs end their decline after the last epoch of our spectropolarimetric data (see e.g. Table \ref{66Oph_rit_data}), suggesting the system's mass-loss rate changed and a complete disk-loss phase was avoided.  The intrinsic polarization data (Fig \ref{fig:ip_all_b}) also exhibit a slow decline throughout the 13 year baseline of the HPOL data, supporting the interpretation of a gradual decline in the disk.

\subsection{Intrinsic Polarization Color Diagrams} \label{intpol}

\citet{dra11} noted the evolution of the ratio of the polarization across the Balmer jump versus the $V$-band polarization occasionally traced out distinctive loops in PCD diagrams. \citet{hau13} explored these diagrams in detail with theoretical models.  Since the polarization change across the Balmer jump is roughly proportional to the density squared whereas the $V$-band polarization is proportional to the density, PCD diagrams offer a useful method to investigate the evolution of Be disks when spectral type and inclination can be constrained \citep{hau13}.  PCD loops observed in 60 Cyg and $\pi$ Aqr were qualitatively modeled using the radiative transfer code HDUST and 1D hydrodynamical 
code SINGLEBE by turning the mass decretion rate on and off.  Subsequent detailed modeling of PCDs demonstrated that the shape, slope, and time-scale of loops depends on spectral type, inclination angle, the base density of the disk, and the temporal behavior of the mass decretion rate
\citep{hau13}.   One important conclusion from these modeling efforts is that steady state disks pause at the top of a PCD loop, while the disk 
is fed by either constant mass-loss or semi-regular mass injections from the star \citep{hau13}.  Our steady disk sample broadly reproduces 
this behavior.  For example, both $\phi$ Per and $\gamma$ Cas (Fig \ref{fig:ip_all}) cluster at high $V$-band polarization 
and high Balmer jump ratios in their PCDs during their steady-state disk stage.\\

However, the following systems have deviations from this steady state disk behavior:  \\

\subsubsection{48 Lib}  48 Lib exhibits clear evidence of a partial PCD loop (Fig \ref{fig:ip_all_a}) in our dataset.   The rise in 48 Lib's intrinsic 
$V$-band polarization (blue points; Fig \ref{fig:ip_all_a}) indicates a disk growth phase, which eventually plateaus into a steady state disk phase.  
Unlike other steady state disks in our sample that sit at the top of the PCD loop during the steady state, 48 Lib's Balmer jump ratio rapidly drops during this phase. Since 48 Lib is a near edge-on inclination \citep{riv06,ste12}, its PCD behavior seems consistent with the modeling scenario outlined in Fig. 9 of \citet{hau13}.  Specifically, these results suggest that 48 Lib was characterized by a high base density ($\approx$ 10$^{-11}$ g cm$^{-3}$) disk and disk orientation which gives a high opacity.   This manifests during the onset of the disk-growth event with a large Balmer jump ratio which later 
decreases at a nearly constant $V$-band polarization as it approaches a steady state phase. Furthermore, it has a steeper decline which is consistent with a later spectral type of B2 to B5 given 48 Lib is B3. \\

\subsubsection{$\psi$ Per} $\psi$ Per experienced a notable rise in intrinsic $V$-band polarization during the 
early 1990's (blue points; Fig \ref{fig:ip_all_b}) indicating a disk growth phase, that seemingly plateaued during the later part of the decade into a stable phase.  $\psi$ Per exhibited a dramatic drop in its Balmer jump polarization as it transitioned to a steady state system.  Given the mid spectral type (B5) and near edge-on inclination 
($\sim$75$^{\circ}$; Table 1) of the system, like 48 Lib, we suggest that the $\psi$ Per PCD behavior could be caused by 
the outer regions of the high inclination, high density disk absorbing significant Balmer jump photons at the onset of the disk growth event \citep{hau13}.  This is further supported by the fact the polarization blueward of the Balmer jump was nearly zero, or essentially optically thick, to polarization of the disk. \\

\subsubsection{28 Cyg} During 28 Cyg's overall long-term decline in disk strength, as measured by its declining intrinsic $V$-band polarization and H$\alpha$ EW (Fig. \ref{fig:ip_all_c}), the system experienced three short, large jumps in 
Balmer jump polarization.  During a 90 day period when the star was monitored regularly, in some cases nightly, the PCD diagram illustrates 3 instances where the Balmer jump ratio spiked to $>$2.5 (compared to the median value of 1.4).  
The separation between these three peaks was 24 and 18 days respectively.  Given its spectral type (B2.5), the Balmer jump polarization is most 
sensitive to the inner regions of the disk when compared to the $V$-band polarization \citep{hau13}. These Balmer jump ratio spikes (that occur at a constant $V$-band polarization) are likely caused by discrete events in the inner disk.  Specifically, we speculate that the spikes stem from stochastic mass injections into the disk. \\

Although outside the scope of this paper, it is clear that detailed modeling of 48 Lib, $\psi$ Per, and 28 Cyg using codes 
such as \citet{ha13a} and \citet{hau13} should be improved and pursued to reproduce the PCD variability observed in these systems.  
Moreover, given the diagnostic potential for systems experiencing full disk growth/disk-loss episodes ($\pi$ Aqr, 60 Cyg; \citealt{dra11}) and episodic disk growth (48 Lib, $\psi$ Per, 28 Cyg), it is certainly clear that enhanced observational 
monitoring of these types of systems should be aggressively pursued.  Without consistent temporal monitoring, it can 
become challenging to constrain the exact time-scale for disk growth and loss events.  Our results for 28 Cyg provide quantitative evidence that even high cadence (i.e. nightly) observations yield interesting PCD phenomenon that could be 
better exploited to diagnose episodic mass injection events. \\

\subsection{One-armed Density Waves} \label{PAvar}
 
\citet{ha13b} presented ad hoc model predictions of the effects of global one-armed oscillations \citep{oka91,oka97} on the time-dependent linear polarization of 
Be disks, including models that used a perturbation pattern characterized by a pattern of opposite overdense and underdense regions \citep{oka97} and models that used a spiral shaped perturbation pattern as adopted by \citet{car09}.  These models (and those of \citealt{car09}) predict that the 
$V$-band polarization and polarization across the Balmer jump should exhibit a clear inclination-dependent modulation with phase (Figs 6-8 of \citealt{ha13b}), although they will be out of phase with one another due to the different radial locations in the disks over which the scattering events occur in each bandpass.  The $V$-band polarization position angle is also predicted to exhibit complex changes with phase (Fig 10; \citealt{ha13b}).  In-spite of these predictions, no conclusive observational evidence of this phenomenon has been reported \citep{car09}.

$\gamma$ Cas is known to exhibit V/R variations indicating the presence of a one-armed density feature in its disk, as seen in its phase-folded He I V/R ratios (Fig \ref{wacky}) compiled from data presented in \citet{mir02}.  Observations were made at the Ritter observatory from 1993 to 2002 of the He I at 5876 \AA, amongst other lines, with a resolving power of $\sim$26000.  The line profile was consistently double peaked and had clear V/R variations  $\gamma$ Cas exhibits a generally stable disk, as diagnosed by its intrinsic polarization (Fig. \ref{fig:ip_all_a}), which indicates its inner disk is being supplied by a generally stable decretion rate. It therefore makes an ideal system to search for long-term polarimetric effects related to its one-armed density feature.  The intrinsic $V$-band polarization, polarization across the Balmer jump, and $V$-band polarization position angle phased to the V/R period of the He I data is shown in 
Fig \ref{wacky}.  The intrinsic $V$-band polarization generally exhibits a double-oscillation pattern over one period that is predicted by \citet{ha13b}.  
The intrinsic polarization across the Balmer jump exhibits variability, albeit no clear indication of the phase-lagged behavior predicted by \citet{ha13b}.  The intrinsic $V$-band polarization position angle suggestively exhibits evidence of cohesive variations as a function of phase, but such variations are an order of magnitude greater than that predicted in the ad hoc models of \citet{ha13b}.  We speculate that one reason these data exhibit a stronger indication of the predicted behavior in $V$-band, rather in the Balmer jump polarization, is that the former is less sensitive to small changes in the inner disk caused by changes in the mass decretion rate.  We encourage future modeling efforts of this phenomenon to 
explore the ramifications of abrupt and gradual changes in the mass decretion rate have on the polarization in disks 
having one-armed density waves.

\subsection{PA Deviations}

\citet{wis10} reported numerous instances of the intrinsic polarization of 60 Cyg and $\pi$ Aqr deviating from their linear trends on a Stokes QU diagram, and found that these deviations in intrinsic polarization position angle were more prominent during large outburst events.  These authors interpreted this 
behavior as either evidence of the injection and subsequent circularization of new blobs of mass into the inner disk region, similar to that noted in \citet{car07}, or as evidence of the injection and subsequent circularization of new blobs at an inclined orbit to the plane of pre-existing disk material.
Many of the Be systems explored in our current study exhibit generally stable, strong disks over most of the duration of our dataset, which suggests we should see analogous evidence of PA deviations in our dataset.     As seen in Fig. \ref{fig:gamCasdev}, $\gamma$ Cas exhibits PA deviance as a function of the intrinsic $V$-band polarization in the system. Analogous figures for our other targets are available in the online-version of this paper.  
	Overall, we do see clear evidence of strong PA deviations in most of our Be stars even though they cannot be correlated to specific outburst events like 60 Cyg and $\pi$ Aqr were. The system that exhibits the smallest level of PA deviations, $\omega$ Ori, is noteworthy as it exhibited evidence of a gradual, albeit incomplete, loss of its disk throughout the time-frame covered by our data.  While errors in the ISP estimate could be inducing errors in the intrinsic PA, we note that the deviations are symmetric about the mean.  If there were errors in our PA determinations, then one would expect the deviations to be systematically offset to one side of the mean.  If there were a PA estimate error then one would expect the deviations systematically offset to one side of the mean.  In general, our results are consistent with systems which have had more precise ISP determinations and time resolved outbursts which suggest the PA deviations are also ``clumpy'' injection events (e.g. 60 Cyg and $\pi$ Aqr).

\subsection{Be+sdO systems}\label{sdO}

Three of our targets are known to have a sub dwarf companion, $\phi$ Per \citep{gies98}, FY CMa \citep{peters08}, and 59 Cyg \citep{peters13}, and all three systems exhibit steady state disks in our data.  This apparent stability could be influenced by the sdO truncating these disks to the maximum 
allowed radius.  Interestingly, we do observe significant variability in the polarization across the Balmer jump in both FY CMa
and 59 Cyg (Fig \ref{fig:ip_all_c}).  We remind the reader that the ISP for 59 Cyg was poorly constrained; however, we explored different assumed total ISP PA values (about PA = 0) and still observed the jumps in the polarization across the Balmer jump, suggesting that they are likely real.

As the sdO is likely heating the outer region of the disk nearest its orbital position, we speculate that this outside heating source may inflate the scale height of this region of the disk.  Given the short periods of order 29-37 days \citep{peters13}, we speculate that when the inflated scale heights of these moderately high inclination disks (Table 1) pass in front of our line of sight with the star, they cause the observed changes in polarization across the 
Balmer jump.   We further speculate that the amount of disk material heated and inflated by $\phi$ Per's more distant (127 day period) sdO companion 
may contribute to why this system exhibits no analogous Balmer jump polarization jumps.  Higher cadence observations that sufficiently sample the suggested puffed-up outer disk over the orbital period of these systems' sdO companion would help to further establish our interpretation of this 
observational phenomenon.

\subsection{Maximum Polarization}

The maximum polarization for a viscous disk in equilibrium is found to be correlated with inclination, spectral type, and base density \citep{hau13,ha13b}.  In most cases the disks in this sample reach some form of steady state.  We then assume that the maximum observed polarization at some point in the 15 year survey is the maximum polarization of a disk around these stars.  Since the effective temperature and inclination can be determined by other means (e.g. spectroscopy and interferometry), we use literature values to then derive the range of base densities for this sample of Be stars (See Table \ref{star_sum}).  Given the model tracks of \cite{hau13}, we find that the base density for these stars lie within $ 8\times 10^{-11}$ to $ 4 \times 10^{-12}\,\rm g cm^{-3}$ (See Fig.\ref{pmax}).

Several items are key to interpreting the base density estimates extracted from Figure 5.  $\pi$ Aqr, for example, exhibits a lower maximum V-band polarization than the 4.2e-11 g cm${-3}$ base density track compared to other stars having a similar stellar effective temperature (Figure 5; right panel) due in part because the system's inclination angle (33.6$^{\circ}$ is lower than that used for the model track (70$^{\circ}$.  Similarly, $\pi$ Aqr  exhibits a higher maximum than expected for the 4.2e-11 g cm${-3}$ base density track (Figure 5; left panel) compared to other systems having a similar inclination due in part to $\pi$ Aqr having a warmer stellar effective temperature than adopted for these models.  Due to this degeneracy, it likely has a similarly large base density as $\phi$ Per around $4.2 \times 10^{-11} \rm g cm^{-3}$.  $\psi$ Per and 59 Cyg have similar maximum polarization and inclination yet have very different effective temperatures.  This then requires a low base density of $8.4 \times 10^{-12} \rm g cm^{-3}$ which exhibits maximum polarization independent of effective temperature.  An object like $\gamma$ Cas, shows behavior similar to that of $\pi$ Aqr in relation to the model tracks. It is likely limited in maximum polarization due to its low inclination rather then its spectral type, so it is consistent with a low base density of $4.2 \times 10^{-11} \rm g cm^{-3}$.  Due to the degeneracy and variable nature of some of the stars observed by HPOL, these limits are not meant to be absolute but rather a first look into the disk properties for the class as a whole.  Each star will require more detailed modeling, but potentially broader statistics could be applied to a wider sample if polarimetry, spectroscopy, and interferometry can be obtained simultaneously to derive the applicable parameters.

\section{Summary}

Motivated by the recent discovery of the diagnostic power of polarization color diagrams (PCD) in characterizing the time-dependent mass decretion 
rate in Be stars \citep{dra11}, and the recent theoretical explorations and predictions of the utility of this diagnostic \citep{ha13a,ha13b,hau13}, we carefully analyzed the long-term polarimetric behavior of 9 additional classical Be stars.  Our targets included systems exhibiting evidence of partial disk-loss/disk-growth episodes as well as systems exhibiting long-term stable disks.  After characterizing and removing the ISP along the line of sight to each of these targets, we found: 

\begin{enumerate}

\item Our steady-state Be disk sample (e.g. $\phi$ Per and $\gamma$ Cas (Fig \ref{fig:ip_all_a}) tend to simultaneously exhibit high intrinsic $V$-band polarization and high Balmer jump ratios in their PCDs, which confirms the behavior predicted by theory \citep{ha13a,hau13}.\\

\item The PCDs for several later spectral type, highly inclined Be stars in our sample (e.g. 48 Lib and $\psi$ Per, Fig \ref{fig:ip_all}) exhibit rapid declines in their Balmer jump ratios at epochs of continuously high intrinsic $V$-band polarizations, which provides the 
first observational confirmation of this predicted \citep{hau13} behavior.  As noted by \citet{hau13}, this phenomenon can be produced when the
base density of the disk is very high, and the outer region of the edge-on disk starts to self absorb a significant number of Balmer jump photons.\\

\item We observe numerous, brief jumps in the intrinsic polarization across the Balmer jump in 28 Cyg, $\phi$ Per, FY CMa, and 59 Cyg.  The high cadence of our 28 Cyg data suggest these jumps are likely caused by stochastic injections of material into the inner disk (of this system).  $\phi$ Per, FY CMa, and 59 Cyg are known sdO systems.  We speculate that the sdO companion in these systems might be thermally inflating the outer region of one side of these disks, and that the observed Balmer jump polarization changes might be related to this inflated disk region passing in front of our line of sight as an alternative to blob injection events.  We recommend future high cadence observations of these systems over the orbital periods of their sdO companions be obtained to confirm our interpretation.\\

\item We observe suggestive evidence of coherent variability in the intrinsic $V$-band polarization and polarization position angle of $\gamma$ Cas that phases with the orbital period of a known one-armed density structure in this disk, similar to the theoretical predictions of \citet{ha13b}.  We speculate that the ``astrophysical noise'' present in these trends, and our non-detection of the predicted \citep{ha13b} phased trend in the polarization across the Balmer jump, could be caused by stochastic changes in the mass decretion rate in this system.\\

\item We also provide constraints on the base densities of quasi-static disks given our sample. Given the degeneracy between inclination and spectral type with the max polarization, we can only estimate the base densities are within $\approx 8\times 10^{-11}$ to $\approx 4 \times 10^{-12}\,\rm g cm^{-3}$.

\end{enumerate}

\section{Acknowledgments}

We would like to thank the referee, Carol Jones, for helpful comments which improved this paper.
We thank Brian Babler for his invaluable assistance with various aspects of HPOL data, and the HPOL and Ritter science teams for support of data acquisition.  We also thank Kenneth H. Nordsieck for providing access to the HPOL spectropolarimeter, and Anatoly Miroshnichenko for supplying an 
electronic copy of his archival $\gamma$ Cas data.
ZHD acknowledges partial support from the University of Washington Pre-MAP program, NSF REU at the University of Toledo, and the Washington Space Grant/NASA Summer Undergraduate Research Program.  HPOL and WUPPE observations were supported 
under NASA contract NAS5-26777 with the University of Wisconsin-Madison.
XH wants to thank FAPESP for the grant 2009/07477-1.
ACC acknowledges support from CNPQ (grant 307076/2012-1) and FAPESP (grant
2010/19029-0).   
Observations at the Ritter Observatory are supported by the NSF under the PREST grant AST 04-40784.  
This study has made use of the SIMBAD database, operated at CDS, Strasbourg, France, the NASA ADS
service, and the STScI MAST Archive. We also like to thank the Astronauts and Space Shuttle support crews for the successful completion of STS-35 and STS-67 which obtained data presented in this paper.

\newpage
\clearpage

\newpage
\begin{table*}
\begin{center}
\footnotesize
\caption{Target Star Summary \label{star_sum}}
\begin{tabular}{lcccccc}
Star & Spectral Type & $T_{eff}$ & Inclination & \# of HPOL Reticon nights & \# of HPOL CCD nights & WUPPE  \\ [1ex] \tableline \\[-1.5ex]
48 Libra & 		B3Ve & 	17645$\pm$554$^{1}$ 	& 		$\sim$90$^{7,8}$ & 			13 & 7 & Astro-2 \\
$\psi$ Persei & B5Ve & 		15767$\pm$509$^{1}$ 	& 		75$\pm$8$^{4}$; 75$^{6}$ & 	9 & 10 & Astro-2 \\
$\phi$ Persei & B2Vpe &		25556$\pm$659$^{1}$ 	& 		84$^{5}$; 72$^{6}$ & 		15 & 11 & Astro-2 \\
28 Cygni & 		B2.5Ve & 	18353$\pm$516$^{1}$		& 		64$^{6}$ & 					4 & 31 & N/A  \\
66 Ophiuchi & 	B2Ve & 		21609$\pm$523$^{1}$ 	& 		48$^{6}$ & 					7 & 3 & N/A  \\
$\gamma$ Casseopia & 	B0.5IVpe& 	26431$\pm$618$^{1}$ & 44$^{5}$; 76$^{6}$ & 		9 & 14 & N/A  \\
$\omega$ Orionis   & 	B3IIIe  & 	20200$^{2}$ & N/A & 						10 & 8 & N/A  \\
FY Canis Majoris   & 	B1II    & 	21750$\pm$655$^{1}$ & 55$^{6}$ & 					9 & 14 & N/A  \\
59 Cygni & 		B1.5Vnne & 	21750$\pm$655$^{1}$ & 		72$^{5}$ 73$^{6}$ & 		10 & 8 & Astro-2 \\
60 Cygni & 		B1Ve & 		27000$^{3}$  & 		N/A & 						2 & 27 & N/A \\
$\pi$ Aquarii & B1Ve & 		26061$\pm$736$^{1}$ & 		33.6$^{6}$ & 				53 & 57 & Astro-1 \\
\end{tabular}
\end{center}
\tablecomments{A summary of the spectropolarimetric data analyzed in this paper.  Spectral types were adopted from SIMBAD, except for 48 Lib which is identified as B3 in a more detailed study \citep{ste12}.  Sources for the effective temperature include $^{1}$\citet{fre05}, $^{2}$\citet{can13}, and $^{3}$\citet{kou00}. Sources for 
the disk inclinations include $^{4}$\citet{del11}, $^{5}$\citet{tou13}, $^{6}$\citet{fre05}, $^{7}$\citet{riv06},  $^{8}$\citet{ste12}.  The number of observations used for each star during the survey is listed by which detector was used. If the the target was observed with the Wisconsin Ultraviolet Photo-Polarimeter Experiment (WUPPE), the mission which took the data is listed.}
\end{table*} 

\newpage
\begin{table*}
\begin{center}
\footnotesize
\caption{ Raw HPOL Data \label{ip_data_raw}}
\begin{tabular}{lccccc}
Target Name & Julian Date & \%Q ($V$-band) & \%U ($V$-band) & \%Err & H$\alpha$ EW\\
[1ex] \tableline \\[-1.5ex]
48 Lib    &   2447679.2 &     -0.614 &     -0.497 &    0.005 &     -21.5 \\
48 Lib    &  2448003.2 &     -0.544 &     -0.591 &    0.003 &      -19.7 \\
48 Lib    & 2448012.2 &     -0.575 &     -0.530 &    0.006 &      -20.0 \\
\end{tabular}
\end{center}
\tablecomments{The raw $V$-band polarization and H$\alpha$ equivalent width is presented for every observation of all of our target 
stars.  The full version of this table is available in the online version from this journal.}
\end{table*}

\newpage
\begin{table*}
\begin{center}
\footnotesize
\caption{Field Star Polarization Data \label{t2}}
\begin{tabular}{lccccccc}
Target Star & Field Star & Spectral Type & Distance(pc) & \%Pol. & \% Error & P.A. & P.A. error \\
[1ex] \tableline \\[-1.5ex]
48 Lib & HD145748 & K5III & 275.4 & 0.420 & 0.035 & 99.6 & 2.4 \\
\nodata & HD144639 & F3IV & 144.5 & 1.060 & 0.048 & 86.7 & 1.3 \\
\nodata & HD145153 & K0III & 120.0 & 0.760 & 0.036 & 86.3 & 1.4 \\
\nodata & HD145897 & K3III & 104.7 & 0.310 & 0.035 & 111.8 & 3.2 \\
\nodata & AVG & \nodata & \nodata  & \nodata & \nodata & 92.93 & 10.73 \\
[1ex] \hline \\[-1.5ex]
$\phi$ Per & \nodata &  \nodata &  \nodata &  \nodata & \nodata  &  \nodata &  \nodata \\
[1ex] \hline \\[-1.5ex]
$\psi$ Per & HD23049 & K4III & 125.9 & 0.500 & 0.032 & 113 & 1.8 \\
\nodata & HD21278 & B5V & 150.0 & 0.480 & 0.120 & 120 & 7.1 \\
\nodata & HD20902 & F5Iab & 162.0 & 0.410 & 0.120 & 100 & 8.3 \\
\nodata & HD21428 & B3V & 168.0 & 0.287 & 0.141 & 127.2 & 13.8 \\
\nodata & HD20809 & B5V & 169.0 & 0.569 & 0.102 & 104 & 5.1 \\
\nodata & AVG & \nodata & \nodata & \nodata & \nodata & 112 & 11.1 \\
[1ex] \hline \\[-1.5ex]
$\pi$ Aqr & HD213119 & K5III & 132 & 0.23 & 0.035 & 110.4 & 4.4\\
\nodata & HD212320 & G6III & 87 & 0.39 & 0.017 & 160.0 & 1.2\\
\nodata & HD211924 & B5III & 251 & 1.17 & 0.037 & 140.3 & 0.9\\
\nodata & HD211838 & B8V & 120 & 0.46 & 0.035 & 138.8 & 2.2\\
\nodata & HD211304 & B9 & 306 & 0.22 & 0.035 & 107.6 & 4.5\\
\nodata & HD211099 & B6 & 417 & 0.26 & 0.038 & 119.2 & 4.2\\
\nodata & AVG & \nodata  & \nodata & \nodata & \nodata & 109.22 & 1.61 \\
[1ex] \hline \\[-1.5ex]
60 Cyg & HD198915 & B6V & 475 & 0.19 & 0.120 & 15.0 & 17.5\\
[1ex] \hline \\[-1.5ex]
$\omega$ Ori & HD36267 & B5V & 143 & 0.133 & 0.026 & 164.7 & 5.6 \\
\nodata & HD37606 & B8V & 242 & 0.150 & 0.032 & 36.0 & 6.1 \\
\nodata & HD38650 & B9 & 352 & 0.134 & 0.023 & 42.0 & 4.9 \\
\nodata & HD38900 & B9 & 360 & 0.074 & 0.026 & 20.0 & 10.0 \\
\nodata & AVG &  \nodata & \nodata & \nodata  & \nodata & 36 & 9.7 \\
[1ex] \hline \\[-1.5ex]
28 Cyg & HD192383 & G5III & 109 & 0.190 & 0.030 & 131.2 & 4.5 \\
\nodata & HD192934 & A0V & 109 & 0.170 & 0.050 & 71.1 & 8.4 \\
\nodata & HD192182 & G8III & 113 & 0.080 & 0.030 & 77.3 & 10.6 \\
\nodata & HD189751 & K1III & 155 & 0.150 & 0.040 & 63.5 & 7.6 \\
\nodata & HD192124 & A5III & 156 & 0.070 & 0.030 & 61.6 & 12.1 \\
\nodata & HD191046 & K0III & 159 & 0.080 & 0.032 & 80 & 11.3 \\
\nodata & HD193636 & A7III & 197 & 0.060 & 0.060 & 3.4 & 26.6 \\
\nodata & HD191045 & K5III & 229 & 0.130 & 0.032 & 61.0 & 7.0 \\
\nodata & HD192745 & A0V & 254 & 0.120 & 0.050 & 86.0 & 11.8 \\
\nodata & HD194357 & B9II & 316 & 0.440 & 0.032 & 76.0 & 2.1 \\
\nodata & HD227421 & A5III & 347 & 0.260 & 0.080 & 94.7 & 8.7 \\
\nodata & AVG &  \nodata &  \nodata &  \nodata &  \nodata & 79 & 35 \\
[1ex] \hline \\[-1.5ex]
66 Oph & HD162177 & A0 & 288 & 0.600 & 0.060 & 77.7 & 2.9 \\
\nodata & HD162954 & B7 & 288 & 0.740 & 0.035 & 75.7 & 1.4 \\
\nodata & HD163592 & B8 & 288 & 0.820 & 0.042 & 92.8 & 1.5 \\
\nodata & HD162649 & A0 & 302 & 0.960 & 0.073 & 79.5 & 2.2 \\
\nodata & HD164097 & A0 & 302 & 0.670 & 0.066 & 68.8 & 2.8 \\
\nodata & HD162993 & A0 & 331 & 0.630 & 0.073 & 73.4 & 3.3 \\
\nodata & HD163152 & A0 & 347 & 0.600 & 0.073 & 75.3 & 3.5 \\
\nodata & HD163591 & B8 & 347 & 1.030 & 0.055 & 85.8 & 1.5 \\
\nodata & AVG &  \nodata & \nodata & \nodata  & \nodata & 81 & 8.5 \\
[1ex] \hline \\[-1.5ex]
FY CMa & HD56094 & B2IV & 1645 & 0.310 & 0.100 & 126.4 & 9.2 \\
\nodata & HD59612 & A5I & 686 & 0.460 & 0.035 & 159.1 & 2.2 \\
\nodata & HD61227 & F0II & 391 & 1.140 & 0.040 & 125.0 & 1.0 \\
\nodata & AVG &  \nodata &  \nodata &  \nodata & \nodata & 136 & 18 \\
[1ex] \hline \\[-1.5ex]
59 Cyg & HD202654 & B2V & 549 & 0.180 & 0.200 & 24.0 & 29.1\\
\nodata & BD+44.3627 & B3V & 504 & 0.780 & 0.180 & 46.0 & 6.6\\
\nodata & HD198915 & B6V & 475 & 0.190 & 0.120 & 15.0 & 17.5\\
\nodata & HD201836 & B5V & 246 & 1.060 & 0.200 & 21.0 & 5.4\\
\nodata & AVG &  \nodata &  \nodata &  \nodata &  \nodata & 29 & 35\\
\end{tabular}
\end{center}
\tablecomments{Archival $V$-band polarization data for field stars used in this manuscript, compiled from  $^{1}$\citet{mat70} and $^{2}$\citet{hei00}, 
are listed.  Spectral types for the field stars were obtained from SIMBAD, and distance measurements were obtained from the 
Hipparcos catalog \citep{per97}.}
\end{table*} 

\newpage
\begin{table*}
\begin{center}
\footnotesize
\caption{Final Interstellar Polarization Parameters \label{isp}}
\begin{tabular}{lccccc}

Star & $\%$P & $\theta$ & $\lambda_{max}$ & $\%$P$_{\perp}$ & $\%$P$_{\parallel}$ \\
[1ex] \tableline \\[-1.5ex]
48 Lib &      0.86 &        93 &        5593 &      -0.78 &      -0.36 \\
$\phi$ Per &  0.76 &        99 &        4539 &       0.40 &      -0.65  \\
$\gamma$ Cas & 0.31 &        95 &        3786 &       0.15 &      -0.27 \\
66 Oph &       0.45 &        81 &        5466 &       0.25 &      -0.37 \\
$\omega$ Ori & 0.23 &        110 &        4442 &       0.12 &      -0.19 \\
$\psi$ Per &    0.09 &        112 &        528 &      0.065 &     -0.06 \\
28 Cyg &     0.25 &        79 &        5496 &      -0.22 &      -0.10 \\
FY CMa &  0.04 &        136 &        3545 &     -0.03&      0.03 \\
59 Cyg &  0.51 &        0 &        5805 &      -0.18 &       0.47\\
60 Cyg &     0.11 &        41.5 &        5349 &      0.08 &      0.07 \\
$\pi$ Aqr &  0.50 &        108.7 &        5123 &      -0.45 &      -0.22 \\
\end{tabular}
\end{center}
\tablecomments{The final parameters for the parallel component ($\%$P$_{\parallel}$), perpendicular component ($\%$P$_{\perp}$), and net ($\%$P) 
interstellar polarization for each of our target stars is compiled. }
\end{table*}

\newpage
\begin{table*}
\begin{center}
\footnotesize
\caption{ Intrinsic Polarization \label{ip_data}}
\begin{tabular}{lccccc}
Target Name & Julian Date & \%Q ($V$-band) & \%U ($V$-band) & \%Err \\
[1ex] \tableline \\[-1.5ex]
48 Lib &       2447679.2 &      0.233 &     -0.410 &    0.005 \\
48 Lib &        2448003.2 &      0.304 &     -0.504 &    0.003 \\
48 Lib &        2448012.2 &      0.273 &     -0.442 &    0.006 \\
\end{tabular}
\end{center}
\tablecomments{The intrinsic $V$-band polarization is presented for every observation of all of our target 
stars.  The full version of this table is available in the online version of this journal.}
\end{table*}

\newpage
\begin{table*}
\begin{center}
\footnotesize
\caption{ 66 Oph Ritter Data \label{66Oph_rit_data}}
\begin{tabular}{lc}
Julian Date &  EW \\
[1ex] \tableline \\[-1.5ex]
2450930.9 &    -38.1\\
2450952.8 &      -36.4\\
2450999.7 &       -35.2\\
2451058.6 &      -34.7\\
2451777.6 &         -26.3\\
2451809.5 &        -25.2\\
2451814.5 &        -25.3\\
2451987.9 &         -23.9\\
2452033.8 &       -23.0\\
2452086.8 &       -22.1\\
2452105.7 &          -22.7\\
2452128.6 &       -22.3\\
2452164.5 &        -21.8\\
2452450.7 &           -20.5\\
2452461.8 &         -20.3\\
2452757.8 &        -15.4\\
2452792.8 &         -14.6\\
2452818.8 &       -14.2\\
2452849.6 &         -15.1\\
2452876.6 &       -14.8\\
2452898.6 &       -15.1\\
2453162.8 &        -12.4\\
2453187.7 &        -12.4\\
2453223.6 &       -12.3\\
2453259.6 &      -12.2\\
2453553.7 &        -10.9\\
2453623.6 &         -12.2\\
2453648.5 &         -11.5\\
\end{tabular}
\end{center}
\tablecomments{Supplementary H$\alpha$ EW measurements from spectra of 66 Oph obtained at Ritter Observatory.}
\end{table*}

\newpage
\clearpage
\begin{subfigures}
\begin{figure*}
	\begin{center}
	\Large
	\textit{\textbf{48 Lib}}\\
	\includegraphics[width=\textwidth,trim = 0mm 0mm 0mm 133mm, clip]{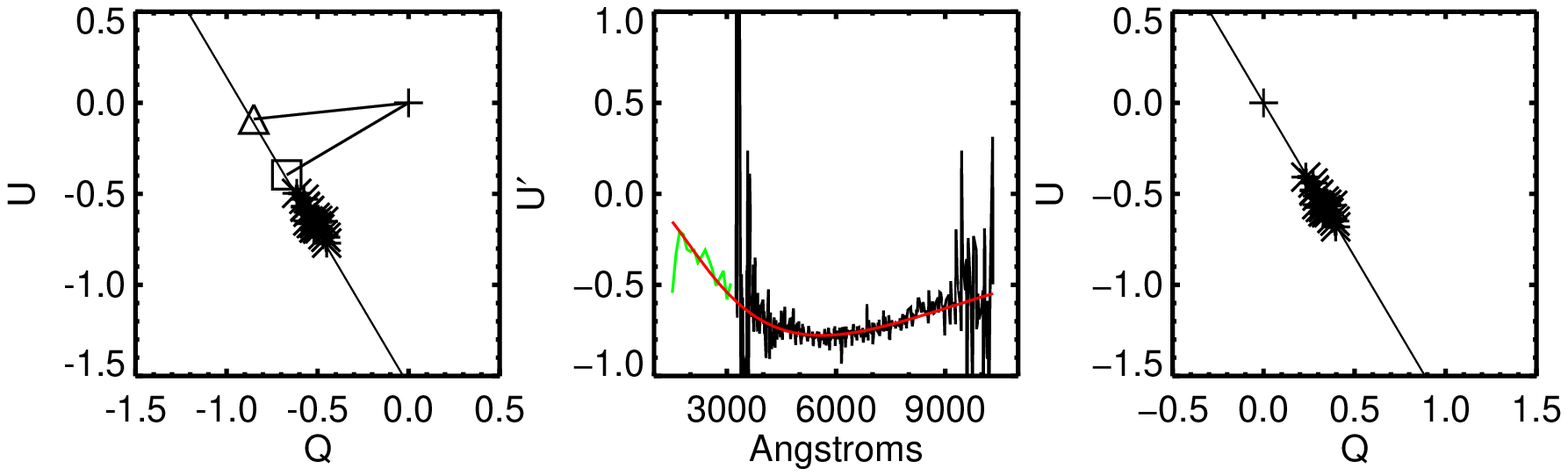}
	\Large
	\textit{\textbf{$\phi$ Per}}\\
	\includegraphics[width=\textwidth,trim = 0mm 0mm 0mm 133mm, clip]{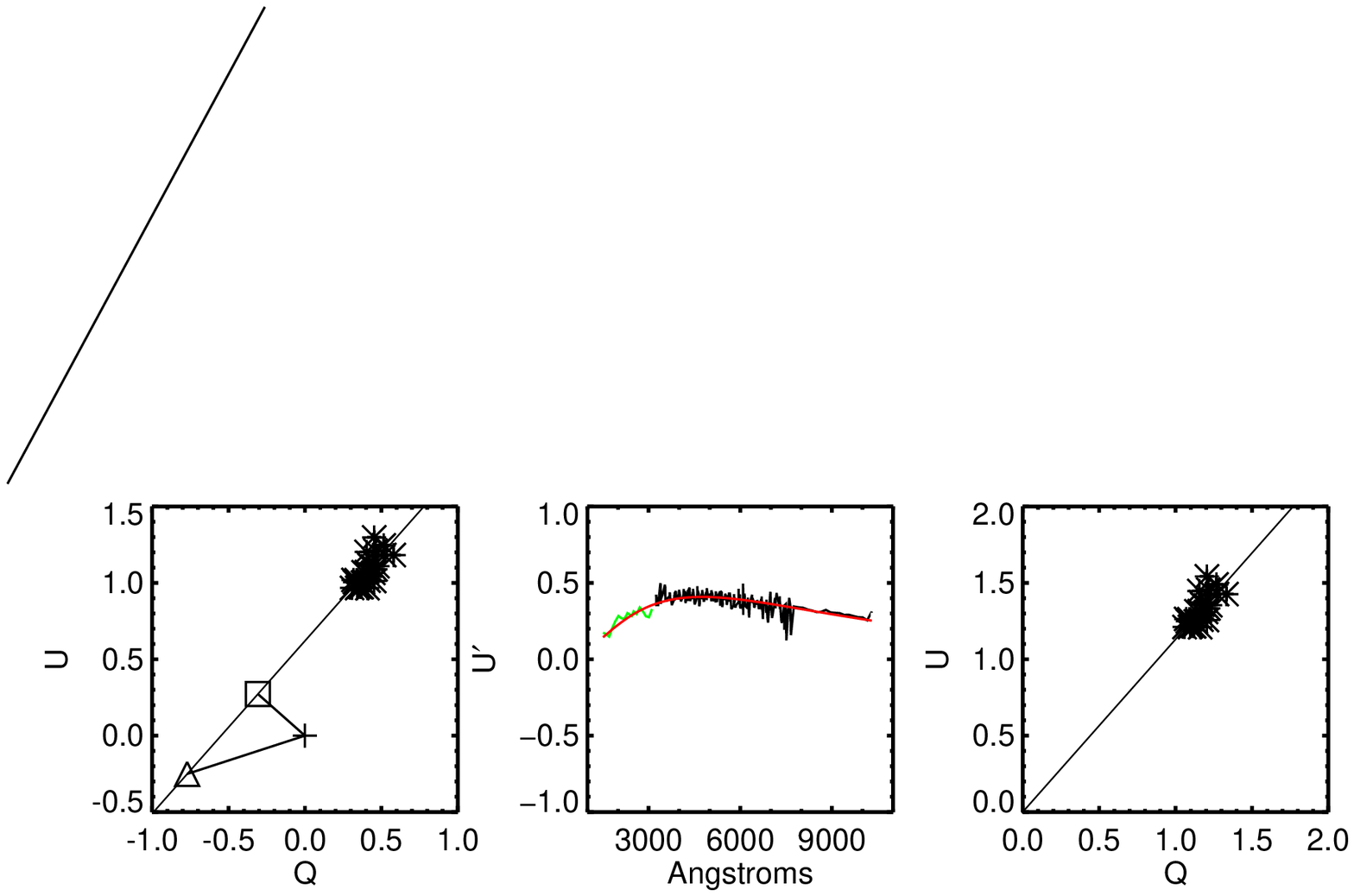}
	\Large
	\textit{\textbf{$\gamma$ Cas}}\\
	\includegraphics[width=\textwidth,trim = 0mm 0mm 0mm 133mm, clip]{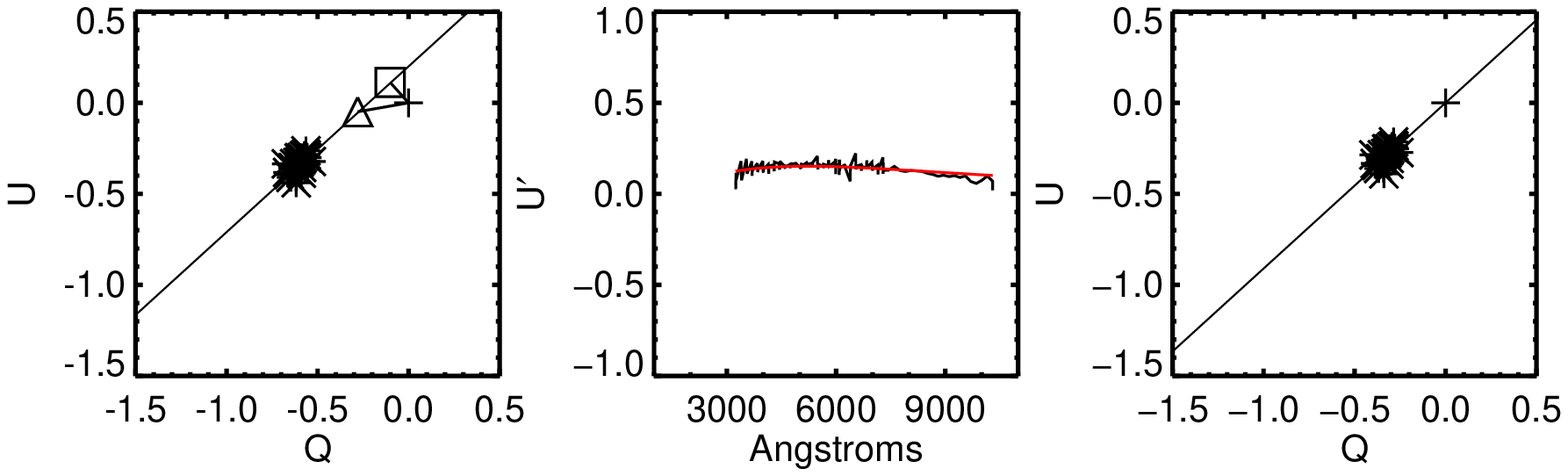}
	\end{center}
	
\caption{\label{fig:isp_all_a}  The left panel depicts the raw $V$-band polarization for a target star plotted on a Stokes QU diagram along with the intrinsic disk position angle vector overlayed.  Either the best fit linear regression or the multi-epoch polarization change across the Balmer jump was used to define the position angle of the disk major axis.  The ISP$_{\perp}$ (square) and ISP$_{total}$ (triangle) vectors are also overlayed.  The center panel depicts the wavelength dependence of the U' data (WUPPE data shown in green, HPOL data shown in black), formed by rotating the $V$-band polarization by the PA of the disk major axis.  The best fit Serkowski law fit (red) parameterizes the ISP$_{\perp}$ and ISP $\lambda_{max}$ components (Table \ref{isp}).}
\end{figure*}

\begin{figure*}
	\begin{center}
	\Large
	\textit{\textbf{66 Oph}}\\
	\includegraphics[width=\textwidth,trim = 0mm 0mm 0mm 133mm, clip]{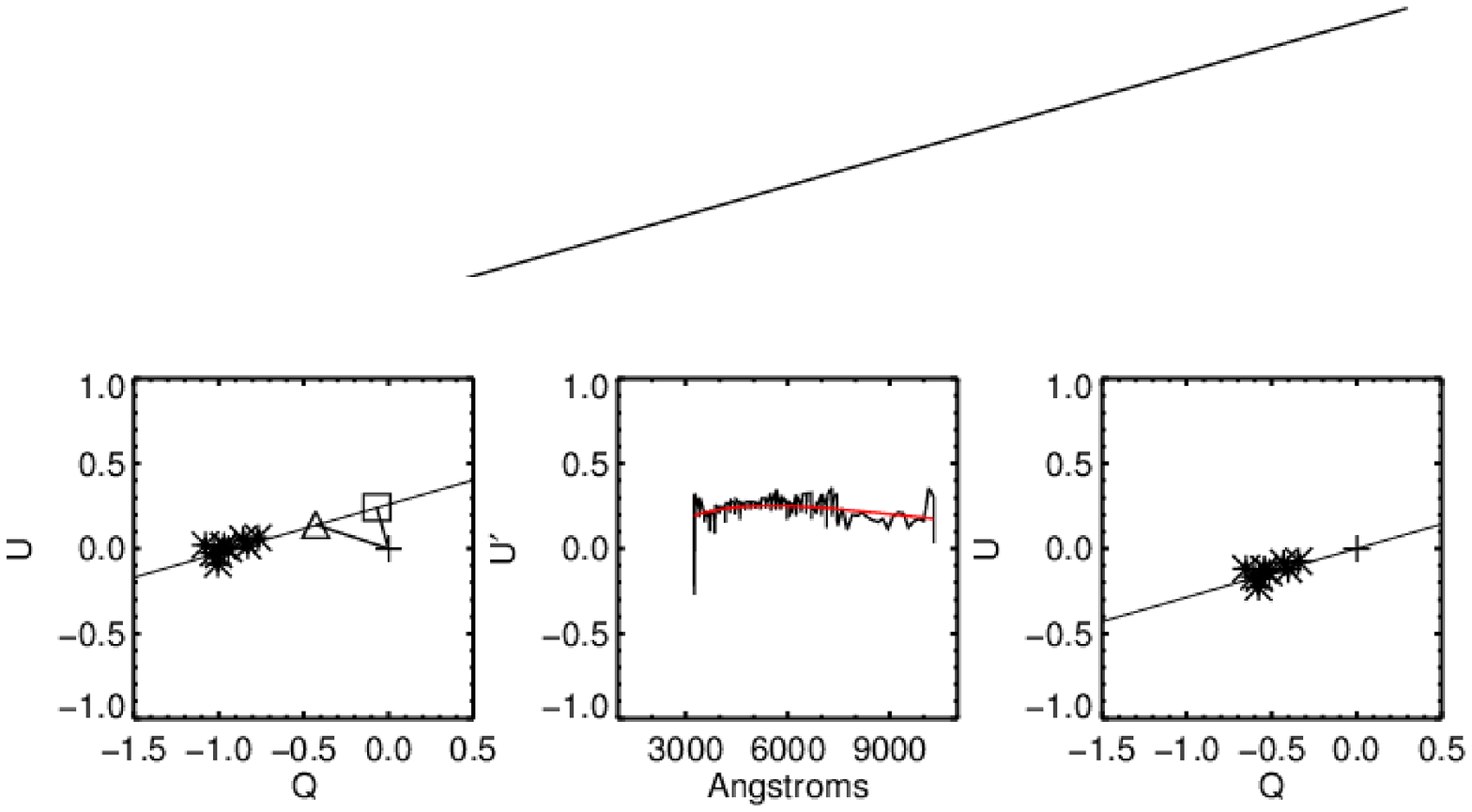}
	\Large
	\textit{\textbf{$\omega$ Ori}}\\
	\includegraphics[width=\textwidth,trim = 0mm 0mm 0mm 133mm, clip]{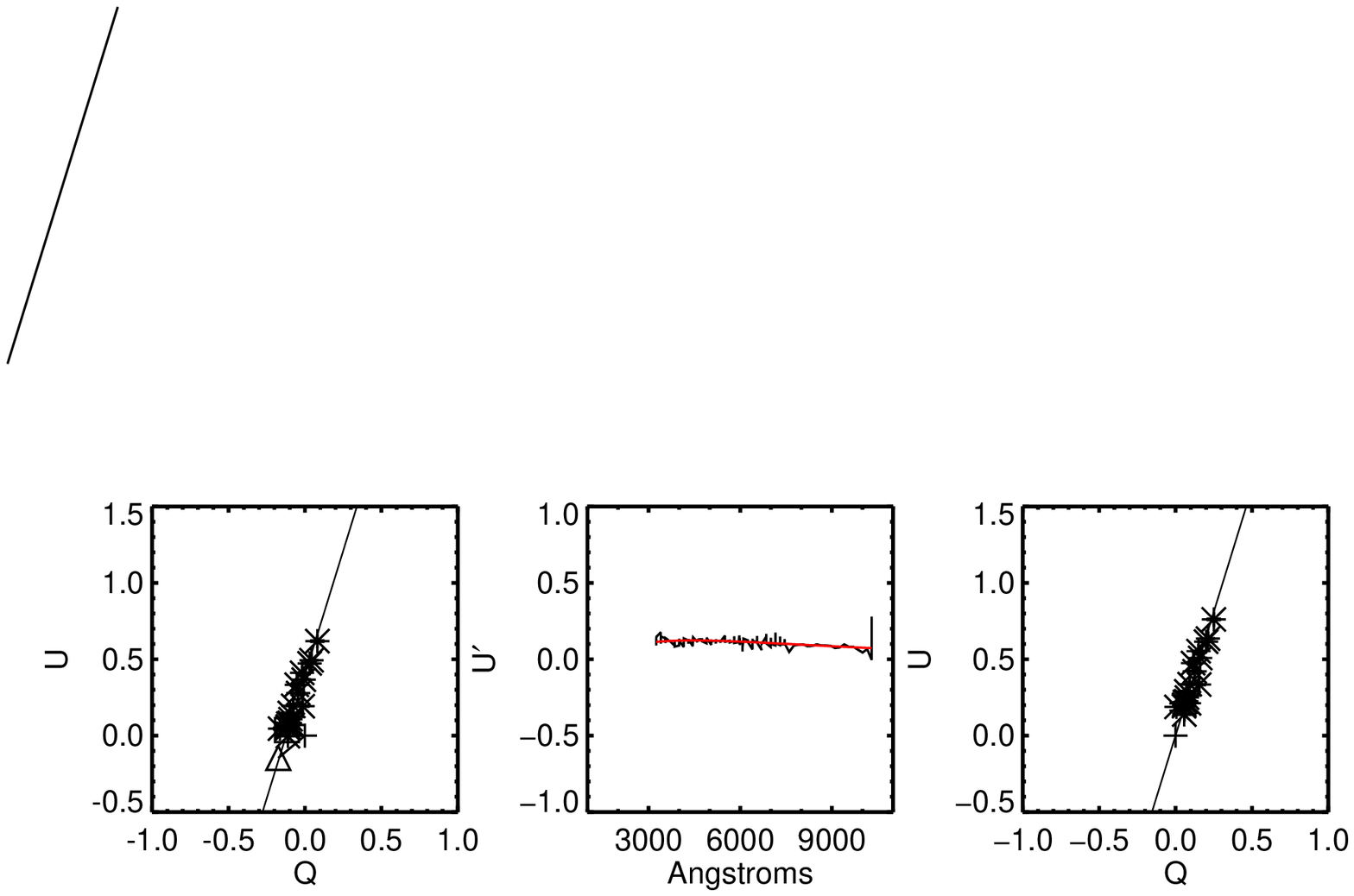}
	\Large
	\textit{\textbf{$\psi$ Per}}\\
	\includegraphics[width=\textwidth,trim = 0mm 0mm 0mm 133mm, clip]{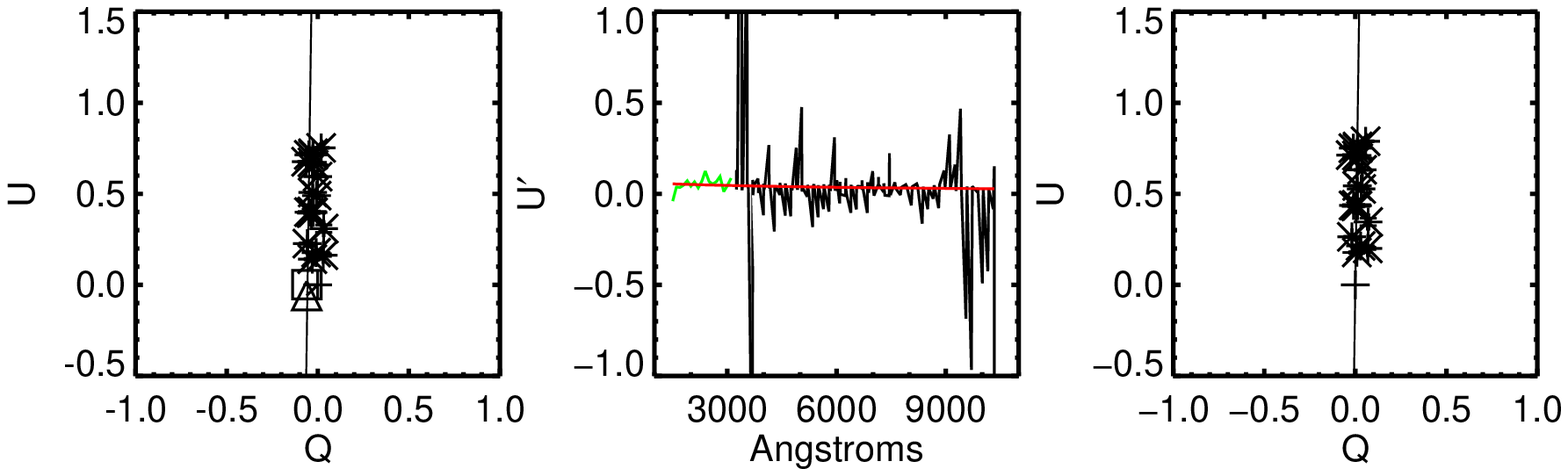}
	\end{center}

\caption{\label{fig:isp_all_b}  The left panel depicts the raw $V$-band polarization for a target star plotted on a Stokes QU diagram along with the intrinsic disk position angle vector overlayed.  Either the best fit linear regression or the multi-epoch polarization change across the Balmer jump was used to define the position angle of the disk major axis.  The ISP$_{\perp}$ (square) and ISP$_{total}$ (triangle) vectors are also overlayed.  The center panel depicts the wavelength dependence of the U' data (WUPPE data shown in green, HPOL data shown in black), formed by rotating the $V$-band polarization by the PA of the disk major axis.  The best fit Serkowski law fit (red) parameterizes the ISP$_{\perp}$ and ISP $\lambda_{max}$ components (Table \ref{isp}).}
\end{figure*}

\begin{figure*}
	\begin{center}
	\Large
	\textit{\textbf{28 Cyg}}\\
	\includegraphics[width=\textwidth,trim = 0mm 0mm 0mm 133mm, clip]{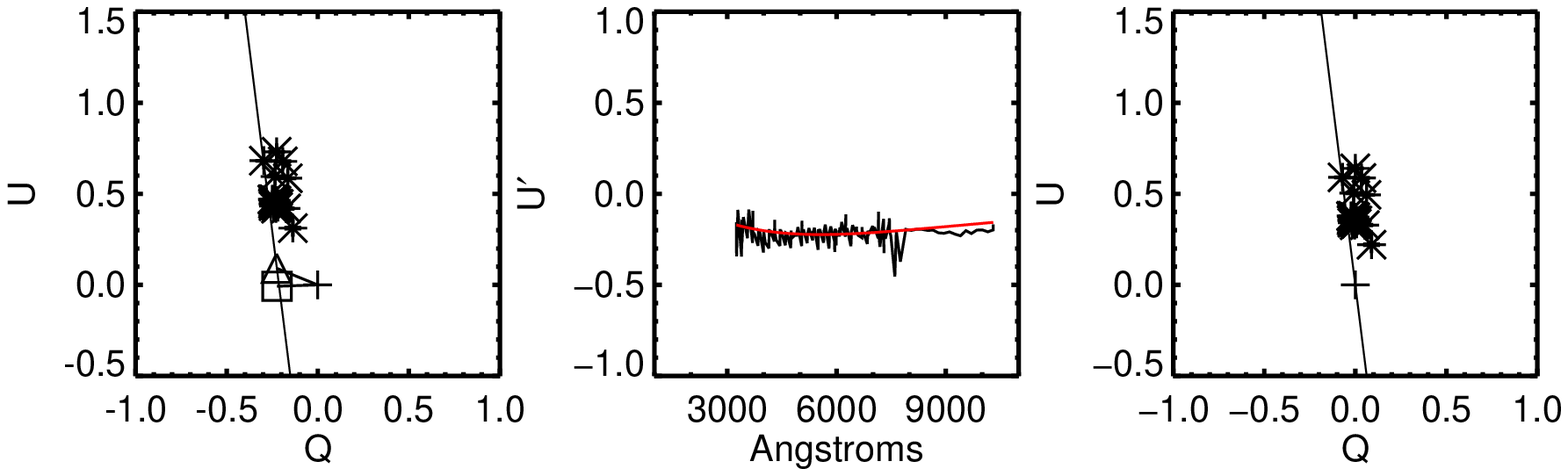}
	\Large
	\textit{\textbf{FY CMa}}\\
	\includegraphics[width=\textwidth,trim = 0mm 0mm 0mm 133mm, clip]{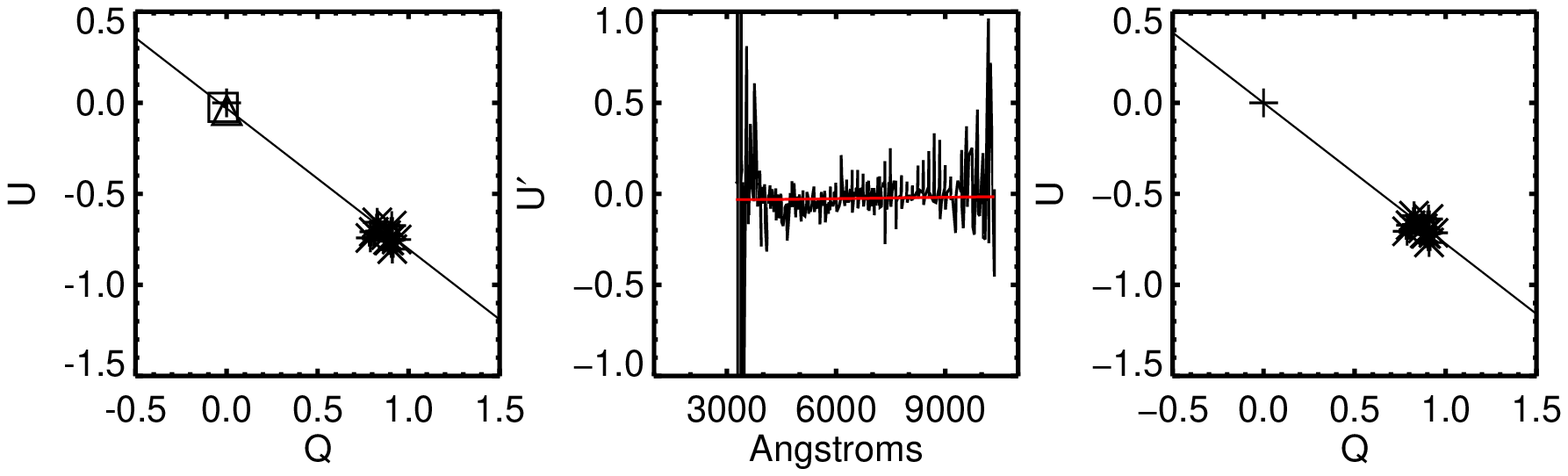}
	\Large
	\textit{\textbf{59 Cyg}}\\
	\includegraphics[width=\textwidth,trim = 0mm 0mm 0mm 133mm, clip]{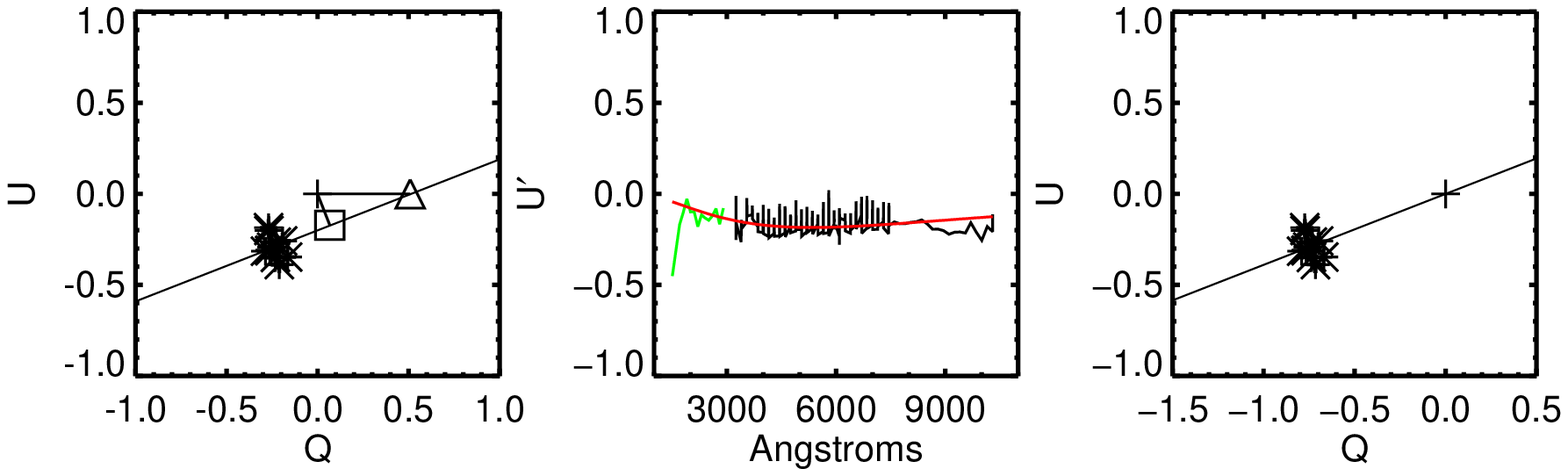}

\caption{\label{fig:isp_all_c}  The left panel depicts the raw $V$-band polarization for a target star plotted on a Stokes QU diagram along with the intrinsic disk position angle vector overlayed.  Either the best fit linear regression or the multi-epoch polarization change across the Balmer jump was used to define the position angle of the disk major axis.  The ISP$_{\perp}$ (square) and ISP$_{total}$ (triangle) vectors are also overlayed.  The center panel depicts the wavelength dependence of the U' data (WUPPE data shown in green, HPOL data shown in black), formed by rotating the $V$-band polarization by the PA of the disk major axis.  The best fit Serkowski law fit (red) parameterizes the ISP$_{\perp}$ and ISP $\lambda_{max}$ components (Table \ref{isp}).}
\end{center}
\label{fig:isp_all}
\end{figure*}
\end{subfigures}

\newpage
\begin{subfigures}
\begin{figure*}
	\begin{center}
	\Large
	\textit{\textbf{48 Lib}}\\
	\includegraphics[width=\textwidth,trim = 0mm 0mm 0mm 133mm, clip]{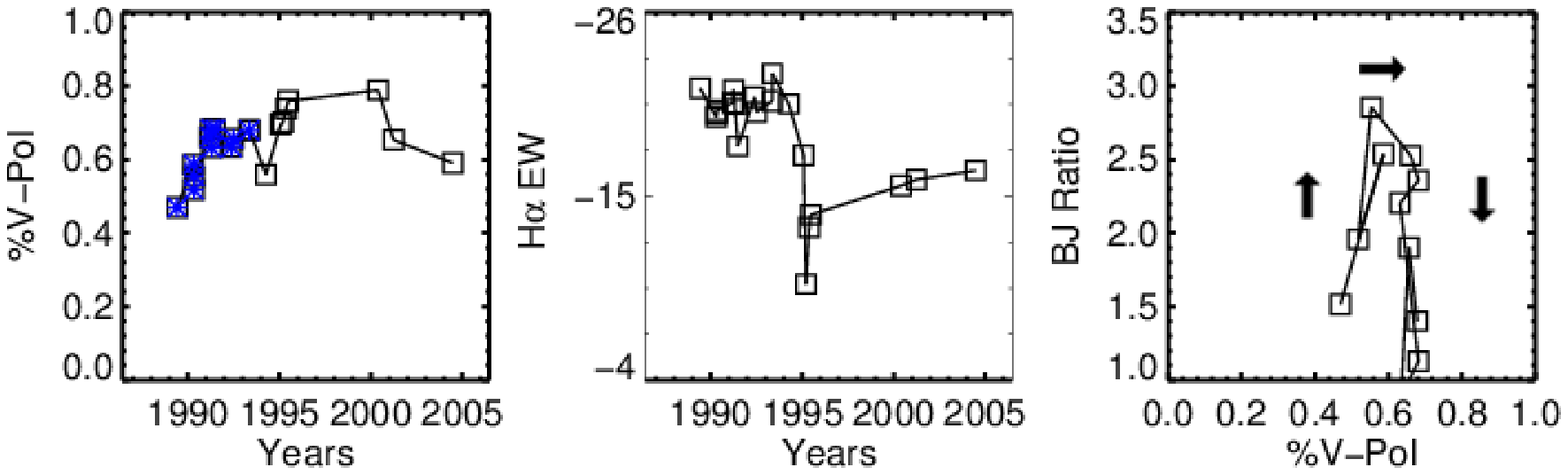}
	\Large
	\textit{\textbf{$\gamma$ Cas}}\\
	\includegraphics[width=\textwidth,trim = 0mm 0mm 0mm 133mm, clip]{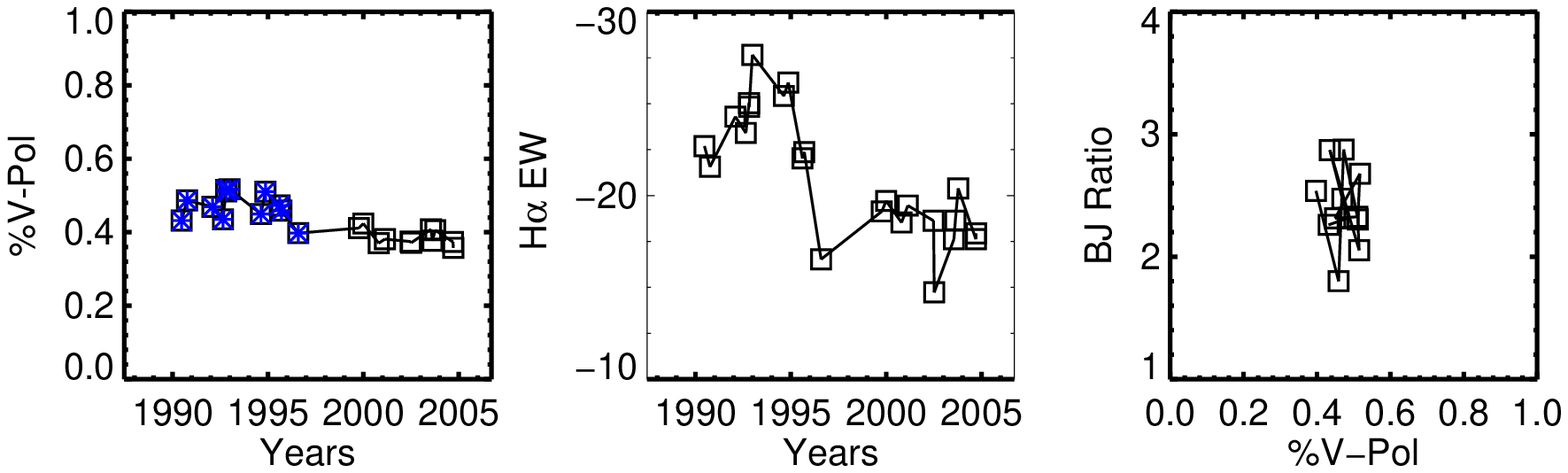}
	\Large
	\textit{\textbf{$\phi$ Per}}\\
	\includegraphics[width=\textwidth,trim = 0mm 0mm 0mm 133mm, clip]{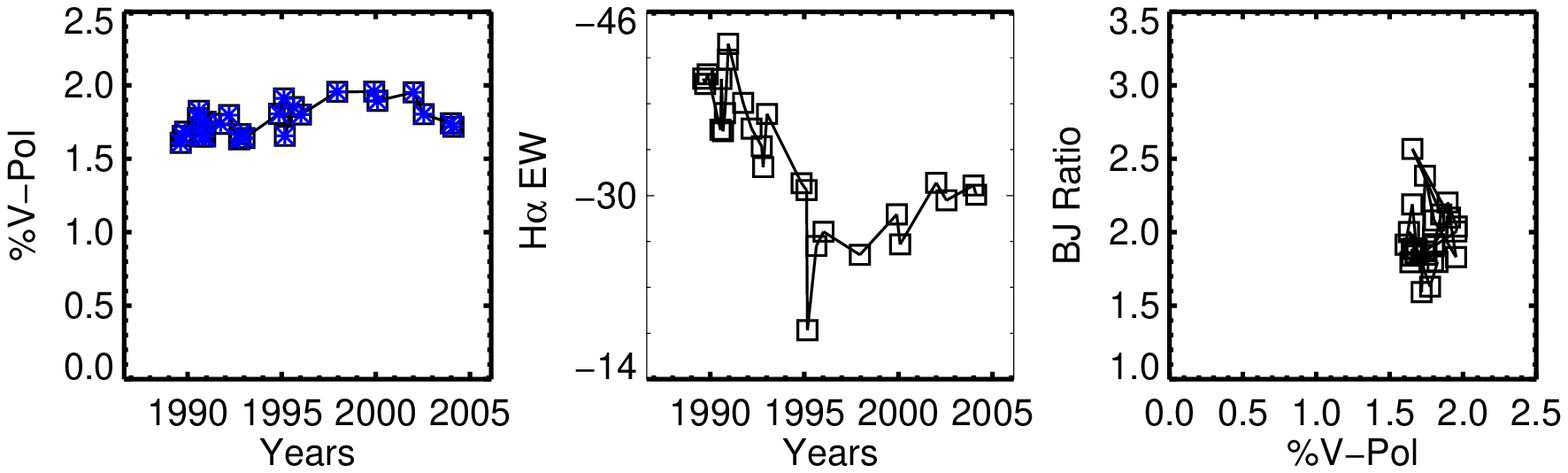}
	\end{center}
	
\caption{\label{fig:ip_all_a} The time evolution of the intrinsic $V$-band polarization of the target stars are in the left panel and the H$\alpha$ equivalent width (EW) are shown in the center panel. EWs measured by HPOL data are shown in squares while EWs measured from spectra obtained at Ritter Observatory are shown in triangles.  The flux offset is due to HPOL not being flux calibrated.  The right panel depicts the PCD for the portion of the $V$-band \% Pol data cross-marked in blue on the left panel.}
\end{figure*}

\begin{figure*}
	\begin{center}
	\Large
	\textit{\textbf{66 Oph}}\\
	\includegraphics[width=\textwidth,trim = 0mm 0mm 0mm 133mm, clip]{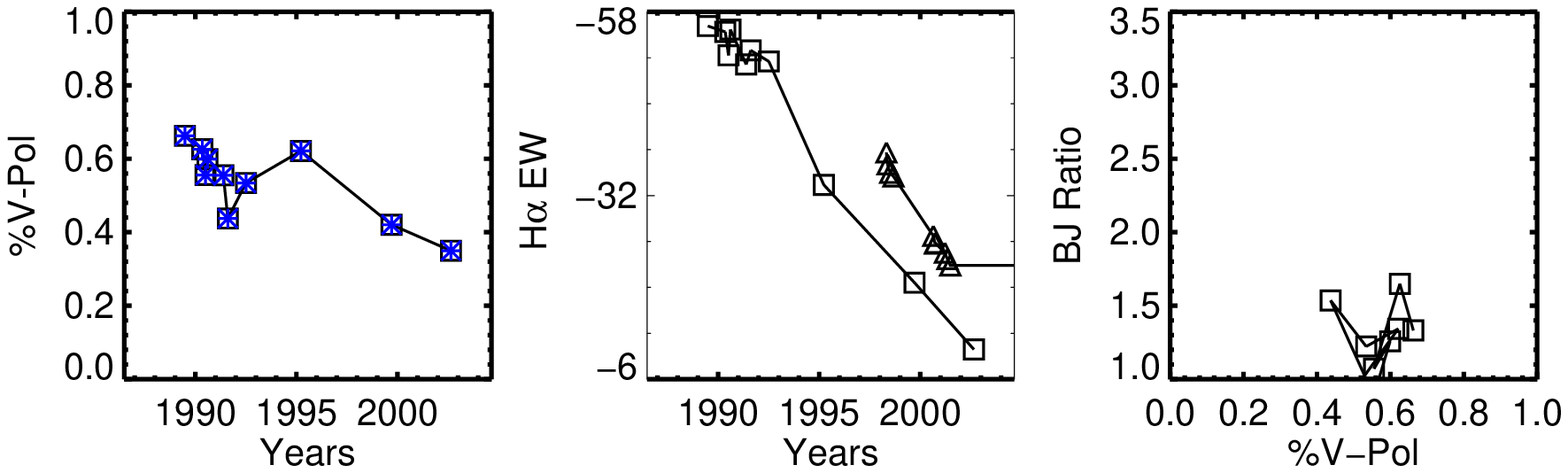}
	\Large
	\textit{\textbf{$\omega$ Ori}}\\
	\includegraphics[width=\textwidth,trim = 0mm 0mm 0mm 133mm, clip]{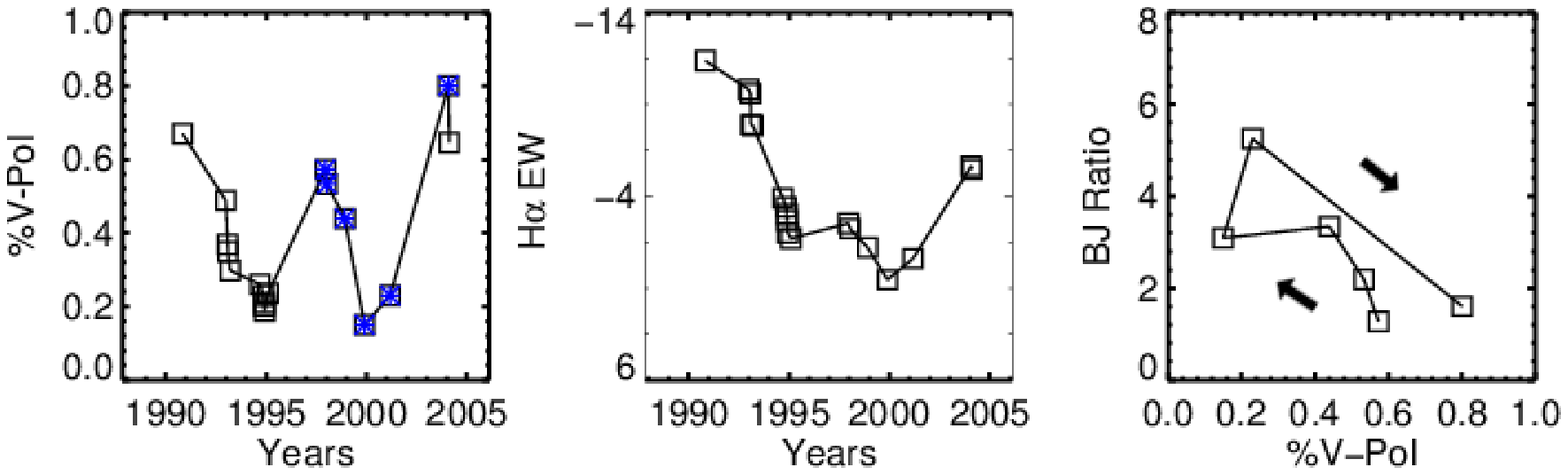}
	\Large
	\textit{\textbf{$\psi$ Per}}\\
	\includegraphics[width=\textwidth,trim = 0mm 0mm 0mm 133mm, clip]{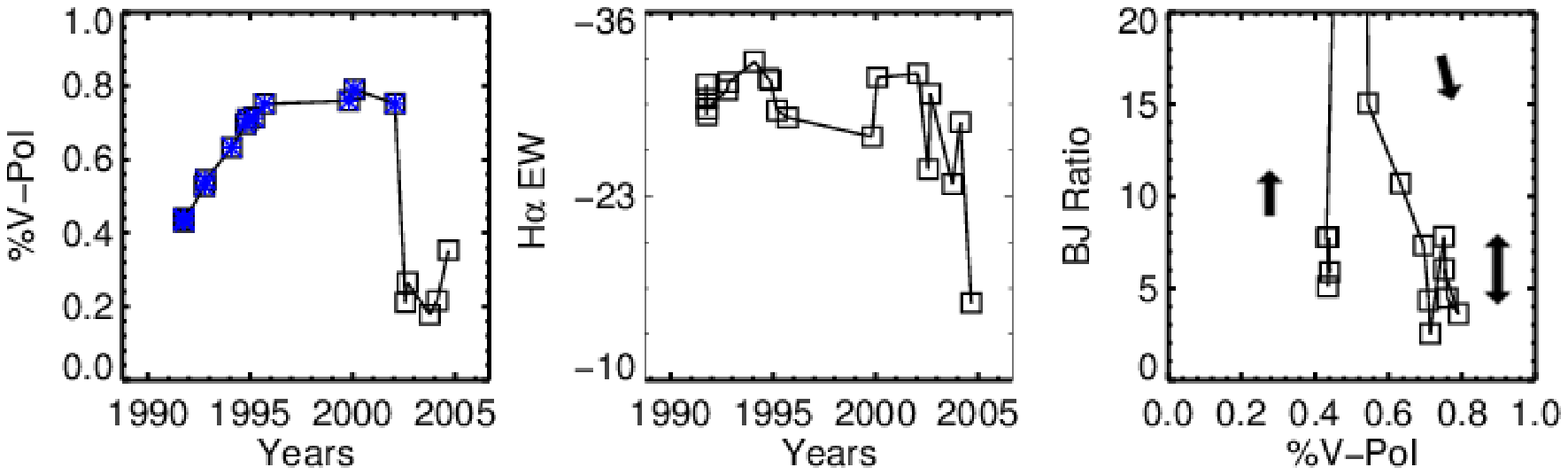}
	\end{center}
	
\caption{\label{fig:ip_all_b} The time evolution of the intrinsic $V$-band polarization of the target stars are in the left panel and the H$\alpha$ equivalent width (EW) are shown in the center panel. EWs measured by HPOL data are shown in squares while EWs measured from spectra obtained at Ritter Observatory are shown in triangles.  The flux offset is due to HPOL not being flux calibrated.  The right panel depicts the PCD for the portion of the $V$-band \% Pol data cross-marked in blue on the left panel.}
\end{figure*}
	
\begin{figure*}
	\begin{center}
	\Large
	\textit{\textbf{28 Cyg}}\\
	\includegraphics[width=\textwidth,trim = 0mm 0mm 0mm 133mm, clip]{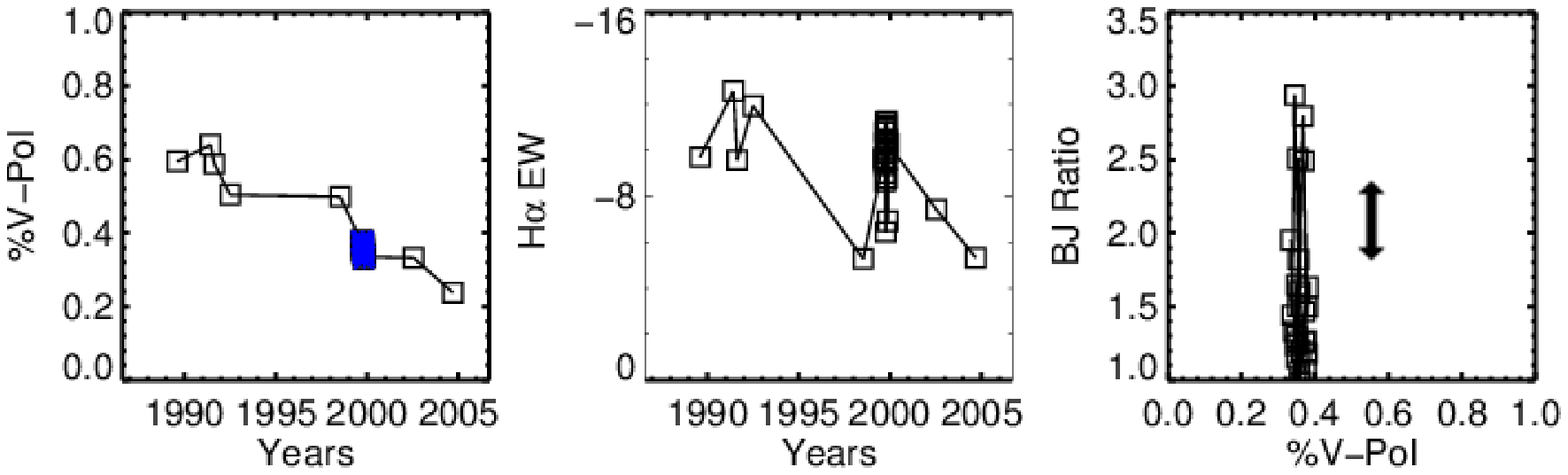}
	\Large
	\textit{\textbf{FY CMa}}\\
	\includegraphics[width=\textwidth,trim = 0mm 0mm 0mm 133mm, clip]{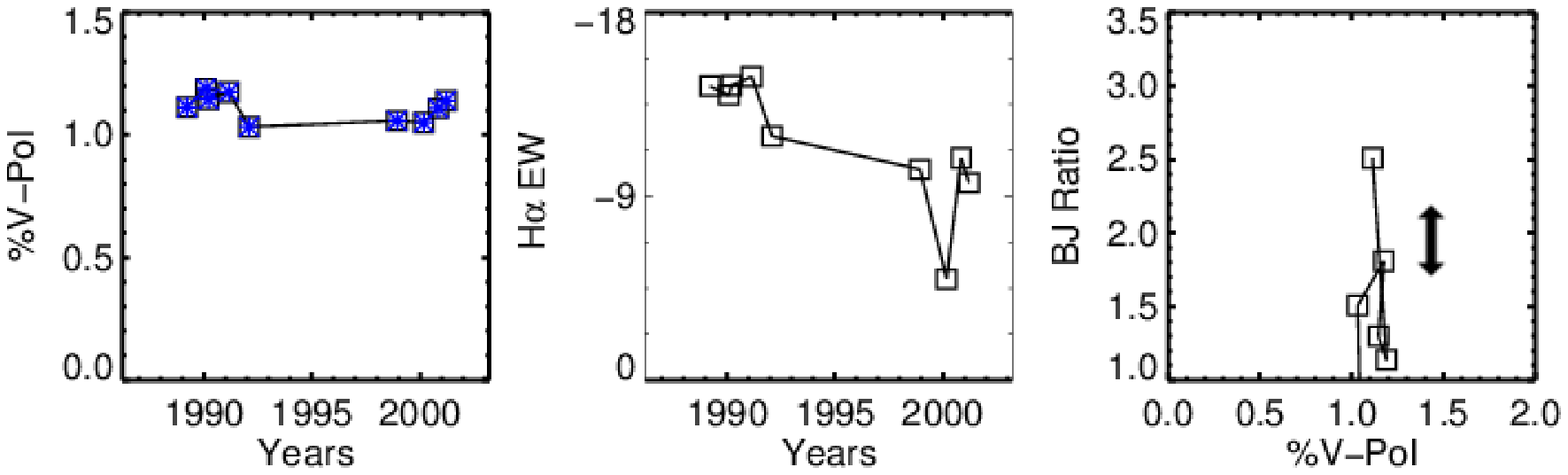}
	\Large
	\textit{\textbf{59 Cyg}}\\
	\includegraphics[width=\textwidth,trim = 0mm 0mm 0mm 133mm, clip]{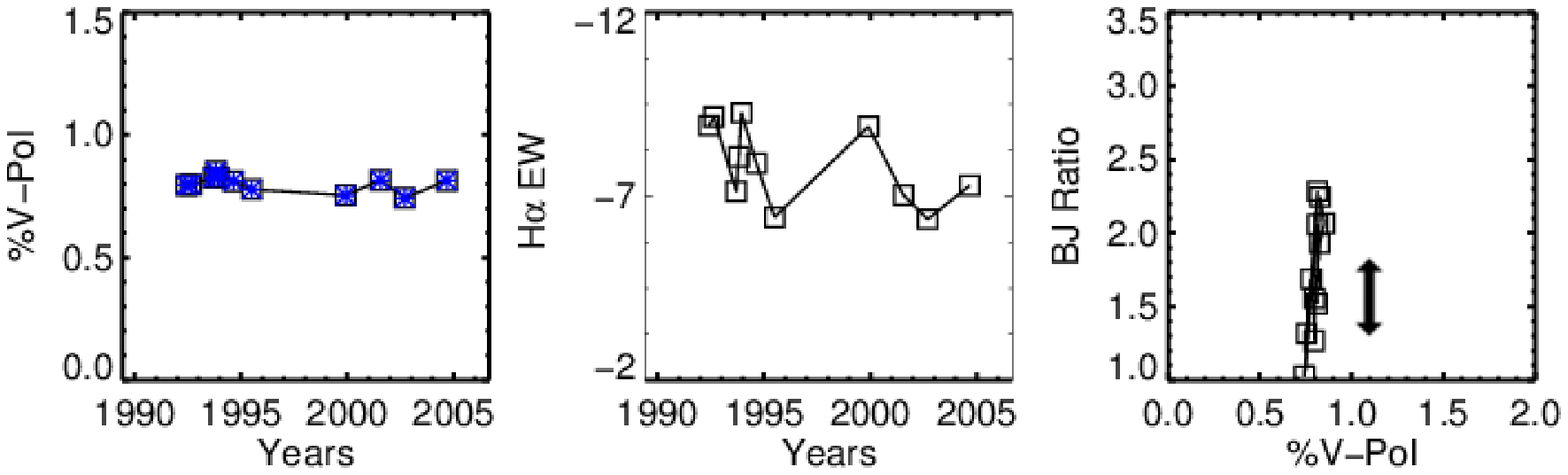}

\caption{\label{fig:ip_all_c} The time evolution of the intrinsic $V$-band polarization of the target stars are in the left panel and the H$\alpha$ equivalent width (EW) are shown in the center panel. EWs measured by HPOL data are shown in squares while EWs measured from spectra obtained at Ritter Observatory are shown in triangles.  The flux offset is due to HPOL not being flux calibrated.  The right panel depicts the PCD for the portion of the $V$-band \% Pol data cross-marked in blue on the left panel.}
\end{center}
\label{fig:ip_all}
\end{figure*}
\end{subfigures}








\newpage
\begin{figure*}
\centering
\includegraphics[trim = 10mm 0mm 0mm 0mm]{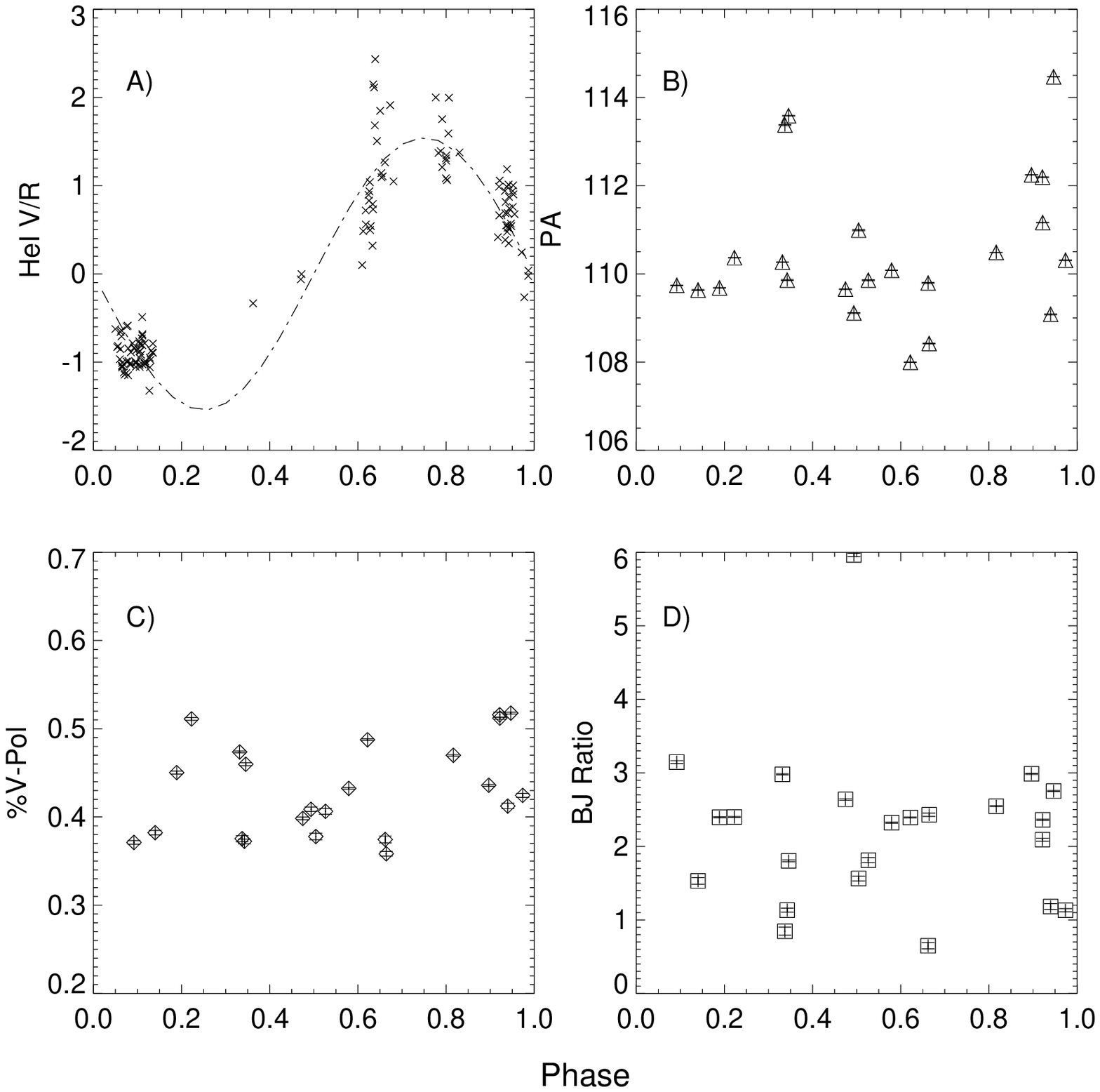}
\caption{The phase folded He I V/R measurements (adopted from \citealt{mir02}; panel A), the phase folded intrinsic $V$-band polarization position angle (panel B), the phase folded intrinsic $V$-band polarization (panel C), and the phase folded intrinsic polarization across the Balmer jump (panel D) is shown for $\gamma$ Cas. \label{wacky}}
\end{figure*}



\newpage
\begin{figure*}
\centering
\includegraphics[]{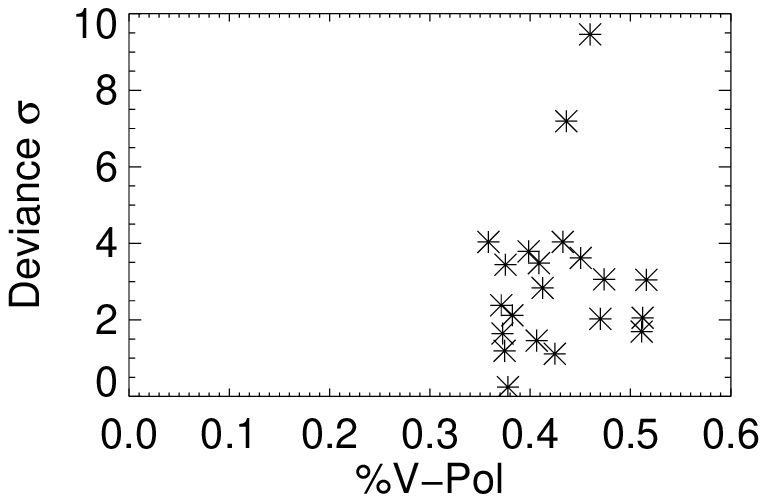}
\caption{Absolute value of the error weighted deviation (deviance) of every $V$-band polarization of $\gamma$ Cas from the best fit line defining its intrinsic disk position angle in its QU diagram (Figure \ref{fig:isp_all}) is plotted as a function of the magnitude of intrinsic $V$-band polarization present in each observation.  The sporadic large deviations mimic those reported by \citet{wis10}.  The online version of this manuscript includes corresponding figures for every star in our target list. \label{fig:gamCasdev} }
\end{figure*}

\newpage
\begin{figure*}
\centering
\includegraphics[]{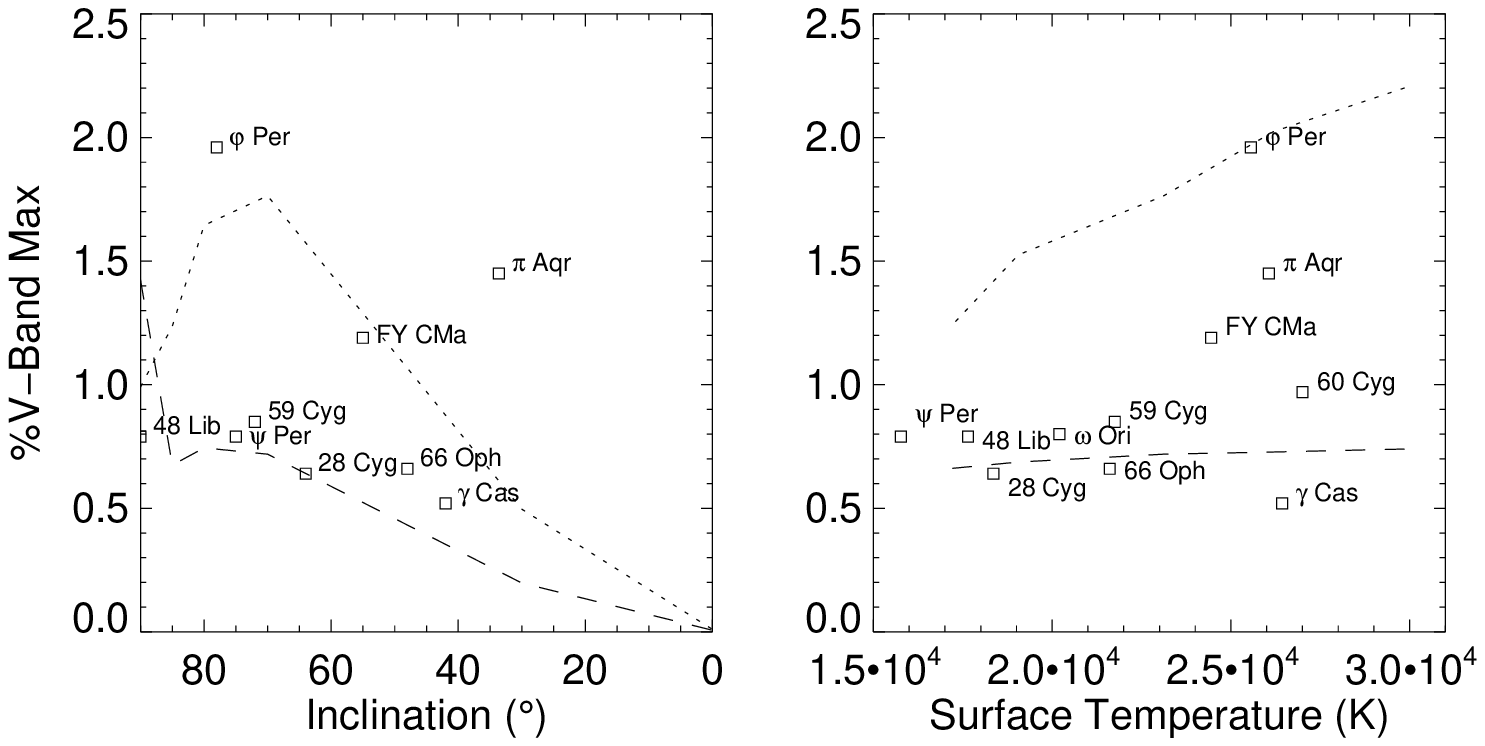}
\caption{ The maximum observed polarization of the target sample is plotted vs disk inclination, on the left, and effective temperature, on the right, when available. Model tracks are shown for two base densities of $8.4*10^{-12}$ on the dashed line and $4.2 \times 10^{-11}$ on the dotted line in $\rm g cm^{-3}$ from \citet{hau13}. The model inclination tracks are of a fixed B2 type star ($T_{\rm eff}$ $\approx$ 23000K). Inclination causes polarization to peak at $i=70-80^{\circ}$. The model effective temperature tracks are from a fixed inclination of $70^{\circ}$. For low base densities, the effect on maximum polarization by effective temperature is constant where as high density models have a growing linear with higher effective temperature.  The polarization maximum is degenerate between the effective temperature and inclination but general limits can be placed on the sample as a whole. \label{pmax}}
\end{figure*}

\end{document}